\documentclass[prb,reprint,superscriptaddress,citeautoscript]{revtex4-1}
\usepackage[fleqn]{amsmath}
\usepackage{graphicx}
\usepackage[para]{threeparttable}
\usepackage{hyperref}
\usepackage[english]{babel}
\usepackage{array}
\usepackage{dcolumn}
\usepackage{multirow}

\begin{document}
\title{First-principles analysis of electron transport in BaSnO$_3$}

\author{Karthik Krishnaswamy}
\affiliation{Department of Electrical and Computer Engineering, University of California, Santa Barbara, CA 93106-9560, USA}
\author{Burak Himmetoglu}
\altaffiliation[Current address: ]{Center for Scientific Computing, California NanoSystems Institute, University of California, Santa Barbara, CA 93106, USA}
\affiliation{Materials Department, University of California, Santa Barbara, CA 93106-5050, USA}
\author{Youngho Kang}
\affiliation{Materials Department, University of California, Santa Barbara, CA 93106-5050, USA}
\author{Anderson Janotti}
\altaffiliation[Current address: ]{Materials Science and Engineering, University of Delaware, Newark, DE 19716-3106, USA}
\affiliation{Materials Department, University of California, Santa Barbara, CA 93106-5050, USA}
\author{Chris G. Van de Walle}
\affiliation{Materials Department, University of California, Santa Barbara, CA 93106-5050, USA}

\begin{abstract}
BaSnO$_3$ (BSO) is a promising transparent conducting oxide (TCO) with reported room-temperature (RT) Hall mobility exceeding 320 cm$^{2}$V$^{-1}$s$^{-1}$.
Among perovskite oxides, it has the highest RT mobility, about 30 times higher than that of the prototypical SrTiO$_3$.
Using first-principles calculations based on hybrid density functional theory, we elucidate the physical mechanisms that govern the mobility by studying the details of LO-phonon and ionized impurity scattering.
A careful numerical analysis to obtain converged results within the relaxation-time approximation of Boltzmann transport theory is presented.
The ${\bf k}$ dependence of the relaxation time is fully taken into account.
We find that the high RT mobility in BSO originates not only from a small effective mass, but also from a significant reduction in the phonon scattering rate compared to other perovskite oxides; the origins of this reduction are identified.
Ionized impurity scattering influences the total mobility even at RT for dopant densities larger than $5\times10^{18}$ cm$^{-3}$, and becomes comparable to LO-phonon scattering for $1\times10^{20}$ cm$^{-3}$ doping, reducing the drift mobility from its intrinsic LO-phonon-limited value of $\sim$594 cm$^{2}$V$^{-1}$s$^{-1}$ to less than 310 cm$^{2}$V$^{-1}$s$^{-1}$.
We suggest pathways to avoid impurity scattering via modulation doping or polar discontinuity doping.
We also explicitly calculate the Hall factor and Hall mobility, allowing a direct comparison to experimental reports for bulk and thin films and providing insights into the nature of the dominant mechanisms that limit mobility in state-of-the art samples.
\end{abstract}

\maketitle

\section{Introduction}
Recent demonstrations\cite{Kim2012,kim-bso-mobility-2,Kim2013a} of electron mobilities as high as 320 cm$^{2}$V$^{-1}$s$^{-1}$ at room temperature (RT) have sparked interest in the transparent cubic perovskite oxide BaSnO$_3$ (BSO) for electronic applications.
The ease of achieving high levels of $n$-type doping\cite{scanlon-bso,Kim2012} (5$\times10^{20}$ cm$^{-3}$) with conductivities on the order of 10$^6$ S~m$^{-1}$ makes it attractive as a transparent conducting oxide (TCO).
Moreover, it has the highest RT mobility among TCOs\cite{tco-rev}.
Its RT mobility is more than an order of magnitude higher than that of perovskite oxides\cite{Wemple1969} with conduction bands derived from $d$ orbitals, of which SrTiO$_3$ (STO)\cite{Himmetoglu2014,Verma2014a,Mikheev2015,Behnia2015} is a prototypical example.

Efforts to further improve BSO's mobility through growth of high-quality bulk as well as thin films have been undertaken by many groups\cite{Kim2013a,Kim2015,Schlom-Piper2016,Raghavan2016}.
However, the characteristics of BSO that impart such a high mobility, and the fundamental limits on this mobility, are still poorly understood.
The small effective mass has been suggested as the primary cause for the high mobility\cite{Kim2012,mizoguchi-bso-opt-2}.
Here we will show that the mass is not the only reason, and that BSO has a significantly lower scattering rate than, for instance, STO.

In this work, we explore the underlying mechanisms responsible for the high RT mobility by calculating the transport properties using Boltzmann transport theory and first-principles calculations.
We use the relaxation time approximation, but unlike the majority of the electron transport studies that assume a constant relaxation time\cite{Sun2016,Pizzi2014,boltztrap,Slassi2016}, we take the ${\bf k}$ dependence of the relaxation time into account.
As recognized in other material systems\cite{Zhou2016,Li2015,Yates2007}, we will see that there is a significant ${\bf k}$ dependence in the relaxation time.
This has important consequences when analyzing the dependence of mobility on carrier concentration and temperature, and also allows us to calculate Hall mobility ($\mu_\text{H}$), which differs from the drift mobility ($\mu$), for comparing against experimental reports.
We also address technical issues related to numerically computing the scattering rates, as well as the importance of adequate sampling of the band structure in order to obtain converged results.

BSO has a 5-atom unit cell that leads to a total of 15 phonon modes, three of which are polar longitudinal optical (LO) modes\cite{stanislavchuk2012}.
In polar crystals, LO phonons tend to dominate scattering at RT compared to other phonons due to their strong long-range coulomb interaction.
In addition, we need to assess ionized impurity scattering, since large concentrations of dopants are intentionally introduced in order to achieve carrier densities as high as $10^{19}-10^{21}$ cm$^{-3}$.

The paper is organized as follows: In Sec.~\ref{sec:structure}, we discuss our first-principles results for the atomic and electronic structure of BSO.
Section~\ref{sec:transportmethod} presents the methodology as well as the computational implementation for calculating transport properties accounting for LO-phonon and ionized impurity scattering via Boltzmann transport theory.
Section~\ref{sec:results} contains the calculated results and a discussion of scattering mechanisms.
Section~\ref{sec:compexp} addresses calculations of Hall factor and Hall mobility and discusses the comparison to experimental transport measurements on bulk and thin films.
In Sec.~\ref{sec:compareSTO}, we address why BSO's mobility is larger than that of other perovskite oxides, with the goal of guiding the search for other high-mobility materials.
In Section~\ref{sec:enhancemob}, finally, we suggest avenues for enhancing the RT mobility of BSO.

\section {Atomic and Electronic structure} \label{sec:structure}
Our first-principles analysis is based on density functional theory (DFT) calculations.
An accurate description of the electronic structure is essential to obtain reliable results for transport properties; we therefore use
the HSE06 hybrid functional\cite{Heyd2003}, which has been shown to yield accurate band structures for solids \cite{Paier2006,Bjaalie2014,Gruneis2014}.
The calculations were performed using the Vienna {\it Ab initio} Simulation Package (VASP)~\cite{Kresse1996} with projector augmented waves\cite{Blochl1994,Kresse1999}. Sn $d$ states were treated as part of the core; we verified this did not affect the calculated structure and
affected the band gap by less than 0.1 eV.
We used a plane-wave basis with 500 eV cutoff, and the default mixing parameter of 25\% and screening parameter of 0.2~\AA$^{-1}$ for HSE06.
An 8$\times$8$\times$8 $k$-point grid with the Monkhorst-Pack mesh was used for Brillouin-zone integrations.

\begin{figure}[!hc]
  \includegraphics[width=0.5\textwidth]{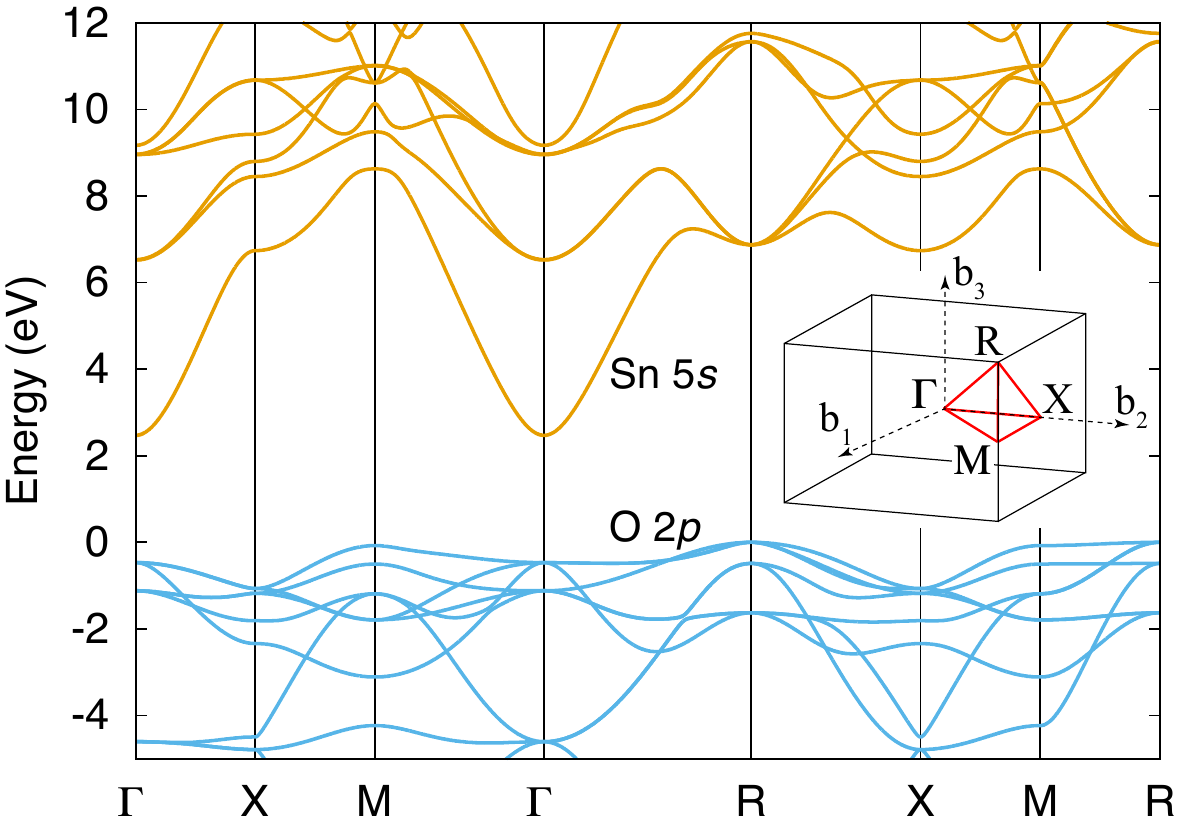}
  \caption{(Color online) Band structure of BSO calculated from first principles using the HSE06 hybrid functional. The inset indicates the high symmetry points in the Brillouin zone.
  The indirect gap $R\rightarrow \Gamma$ is 2.40 eV, and the direct gap at $\Gamma$ is 2.88 eV.}
  \label{fig:bndstruct}
\end{figure}

BSO has a cubic structure with the space group Pm$\bar{3}$m, and has 5 atoms (one Ba, one Sn and 3 O) in its unit cell.
The calculated lattice parameter is 4.13 {\AA}, in good agreement with the experimental value\cite{hinatsu-bso-lat,Kim2013a} of 4.12 \AA.
Our calculated band structure is shown in Fig.~\ref{fig:bndstruct}.
The conduction band (CB) is derived from Sn 5$s$ orbitals and is highly dispersive.
There is also a significant nonparabolicity in the dispersion away from $\Gamma$, which will be quantified in Sec.~\ref{sec:compimp}.
The valence band (VB) is derived from O 2$p$ orbitals, and has much lower dispersion than the CB.

We find an indirect band gap ($\text{R}\rightarrow\Gamma$) of 2.40 eV, in agreement with a previous HSE06 calculation~\cite{scanlon-bso}.
Our calculated direct band gap of 2.88 eV at $\Gamma$ is in reasonable agreement with the reported experimental direct gap of 3.1 eV from optical absorption measurements by Mizoguchi {\it et al.}\cite{mizoguchi-bso-opt-2} and Kim {\it et al.}\cite{Kim2013a}, but disagrees with the value of 3.5 eV reported by Chambers {\it et al.}\cite{Chambers2016} and Li {\it et al}\cite{Li2016}.
Experimentally determined indirect gaps\cite{Chambers2016,Kim2013a} vary between 2.95--3.1 eV and are again larger than our calculated value.
Further work will be needed to resolve the nature and magnitude of the gaps.
We emphasize that our transport calculations do not rely on the value of the band gap, as we will see in the next section.

\section{Methodology for transport calculations} \label{sec:transportmethod}

\subsection{Boltzmann transport theory} \label{sec:boltzmann}

To calculate the electron mobility, we use Boltzmann transport theory within the relaxation time approximation\cite{ziman1960book}.
Due to the cubic symmetry, the conductivity tensor that relates current and the applied field as $j^\alpha\left(\omega\right) = \sigma_{\alpha\beta}\left(\omega\right) E^\beta\left(\omega\right)$ reduces to a scalar\cite{Dresselhaus2008}.
For a given electron density $n$, the drift mobility $\mu$ is obtained from the dc limit ($\omega \rightarrow 0$) of the conductivity $\sigma$ as
\begin{equation}
  \mu = \sigma/ne = 2 \frac{e}{n} \sum_{n} \int_{\Omega} \frac{d^3{k}}{(2\pi)^3}~\tau_{n}\left({\bf k}\right)\left( -\frac{\partial f_{\bf k}}{\partial\varepsilon_{n,\bf k}}\right)\,v_{n,x}^2,
  \label{eq:mu}
\end{equation}
where $f_{\bf k}$ determines the electron occupation given by the Fermi-Dirac distribution for the Fermi level $\varepsilon_\text{F}$,
$\varepsilon_{n,{\bf k}}$ is the energy of the electronic state at wavevector ${\bf k}$ of band index $n$,
$v_{n,x}$ is the $x$-component of the band velocity at wavevector ${\bf k}$,
$\tau_n(\bf k)$ is the scattering time for different scattering processes,
and the integration is over the Brillouin zone (BZ).
The factor ${\partial f_{{\bf k}}}/{\partial \varepsilon_{n,{\bf k}}}$ in the conductivity integral in Eq.~(\ref{eq:mu}) is peaked at the Fermi level and decays to negligible values within a range of $\pm$10$\, k_\text{B} T$, where $T$ is the temperature and $k_\text{B}$ is the Boltzmann constant.
This restricts $\tau_n({\bf k})$ and ${v}_{n,{\bf {k}}}$ entering in the calculation of mobilities to values around the Fermi level $\varepsilon_\text{F} \pm 10$$\, k_\text{B}T$.

A few groups have performed first-principles studies on other materials systems\cite{Ghosh2016,Kaasbjerg2012,Li2013b,Li2015,Yates2007} that explicitly take the ${\bf k}$ dependence of $\tau$ into account; however, the most common assumption in the literature is to take $\tau$ as a constant\cite{Pizzi2014,boltztrap,Slassi2016,Sun2016}, i.e., to approximate $\tau_n(\bf k)$ to be independent of ${\bf k}$.
The value of $\tau$ is often approximated by taking the value at $\Gamma$\cite{boltztrap} or by treating it as a fitting parameter in the analysis of experimental results.
In this work, we explicitly take the ${\bf k}$ dependence of $\tau$ into account; this allows us to check the validity of the constant-$\tau$ approximation, as well as calculate the Hall mobility.

\subsection{LO-phonon scattering} \label{sec:LOphononmethod}

The ${\bf k}$-dependent rate describing an electron-phonon scattering mechanism is obtained from Fermi's golden rule as\cite{ziman1960book,Li2015,Zhou2016,Giustino2016}
\begin{equation}
\begin{aligned}
  &\tau_n^{-1}({\bf k}) = \frac{2 \pi}{\hbar} \sum_{\nu, m} \int_{\Omega} \frac{d^3q}{\left(2\pi\right)^3} \left|{g}_{{\bf q}\nu}({\bf k},n,m)\right|^2 \left(1 - \hat{v}_{n{\bf k}}\cdot\hat{v}_{m{\bf k} + {\bf q}}\right) \\
  & \times \left[ \left(n_{{\bf q}\nu} + f_{m,{\bf k} + {\bf q}} \right)\delta\left( \varepsilon_{m,{\bf k} + {\bf q}} - \varepsilon_{n,{\bf k}} - \hbar \omega_{{\bf q}\nu} \right) \right. \\
  &\left. + \left(1 + n_{{\bf q}\nu} - f_{m,{\bf k} + {\bf q}} \right)\delta\left( \varepsilon_{m,{\bf k} + {\bf q}} - \varepsilon_{n,{\bf k}} + \hbar \omega_{{\bf q}\nu} \right)\right],
\end{aligned}
\label{eq:tauinv}
\end{equation}
where ${g}_{{\bf q}\nu}({\bf k},n,m)$ is the electron-phonon coupling matrix element between states in bands $n$ and $m$, $n_{{\bf q}\nu}$ is the phonon occupation given by the Bose-Einstein (BE) distribution, and $\hbar\omega_{{\bf q}\nu}$ is the energy of phonon mode $\nu$.
The first energy-conserving $\delta$ function (containing the $-\hbar\omega$ term) represents the phonon absorption process, while the second $\delta$ function (containing the $+\hbar\omega$ term) represents the phonon emission process.
The velocity factor, $\left(1 - \hat{v}_{n{\bf k}}\cdot\hat{v}_{m{\bf k} + {\bf q}}\right)$ accounts for the directionality of the current due to the scattered carriers.
BSO has a single nondegenerate CB and therefore only intraband scattering occurs ($m = n = 1$).

As explained in the Introduction, we focus on LO phonons and neglect other phonons in this study.
To compute the electron-phonon (el-ph) coupling matrix element $g_{{ \bf q}\nu}({\bf k})$ we use a generalized Fr\"ohlich model.
Contributions to the polarizability due to different optical branches are explicitly included by using the expression for the coupling matrix elements derived by Toyozawa\cite{Toyozawa1972,Devreese2010} based on the generalized Lyddane-Sachs-Teller relation (in SI units):
\begin{multline}
  \left|g_{{\bf q}\nu}\right|^2 = \frac{q^2}{\left( q^2 + q^2_{\infty,\text{scr}}{F}(\frac{q}{2k_\text{F}})\right)^2}\left( \frac{e^2 \hbar \omega_{\text{L},\nu}}{2\epsilon_\infty\epsilon_0 V_{cell}} \right)\\
  \times\left[ \frac{\prod_{j} \left( 1 - \frac{\omega^2_{T,j}}{\omega^2_{\text{L},\nu}}\right)}{\prod_{j\neq\nu} \left( 1 - \frac{\omega^2_{L,j}}{\omega^2_{\text{L},\nu}}\right)}\right],
  \label{eq:epmatrix}
\end{multline}
where $\epsilon_0$ is the vacuum permittivity, $\epsilon_\infty$ is the electronic part of the dielectric constant, and $q$ is the phonon wavevector.
For a material with a single LO mode Eq.~(\ref{eq:epmatrix}) reduces to the familiar form derived by Fr\"ohlich\cite{Frohlich1954}.
The phonon energy is approximated to be independent of ${\bf q}$ as in the Fr\"ohlich model.
We use the experimental values determined by Stanislavchuk {\it et al.}\cite{stanislavchuk2012}
for the three polar LO (154, 421 and 723 cm$^{-1}$); or 18, 51 and 88 meV)
and the corresponding three doubly-degenerate TO (135, 245 and 628 cm$^{-1}$; or 17, 30 and 78 meV) mode frequencies,
and for the high-frequency dielectric constant ($\epsilon_\infty$=4.3).

In the Fr\"ohlich model\cite{Frohlich1954} as well as in first-principles methods\cite{Giustino2016}, a divergence occurs near $q = 0$ due to the long-range nature of the polarization field in a dielectric.
However, the presence of CB electrons causes this long-range field to be screened, and by including this screening via the screening wavevector $q_{\infty,\text{scr}}$ in the expression for the el-ph matrix element [Eq.~(\ref{eq:epmatrix})], the divergence is avoided\cite{Ehrenreich1959,Kim1978,Giustino2016}.
In principle, screening due to CB electrons is $q$ dependent, and is given by Lindhard theory.
However, in practice, evaluating the full expression from Lindhard theory becomes computational expensive, except at 0 K where an analytic expression can be obtained.\cite{AshcroftMermin1976,Giustino2016}
Therefore, to circumvent the large computational cost, we will compute the screening wavevector $q_{\infty,\text{scr}}$ using Thomas-Fermi theory\cite{AshcroftMermin1976}, which is the $q\rightarrow0$ limit of Lindhard theory, in a medium described by the high-frequency (clamped-ion) dielectric constant $\epsilon_\infty$; the expression will be discussed in Sec.~\ref{sec:compimp}.
The $q$ dependence, which becomes important for phonon wavevectors $q$ comparable to the Fermi wavevector $k_\text{F}$, will be included via the Lindhard function\cite{AshcroftMermin1976,ziman1960book,Giustino2016} $F(q/2k_\text{F})$ in Eq.~(\ref{eq:tauinv}), which has an analytic form only at 0 K:
\begin{equation}
  F(x) = \frac{1}{2} + \frac{1 - x^2}{4x} \ln\left| \frac{1 + x}{1 - x} \right|, x = \frac{q}{2k_\text{F}}.
  \label{eq:lindhard}
\end{equation}
The method to compute $k_\text{F}$ will also be discussed in Sec.~\ref{sec:compimp}.

In principle, the screening also affects the LO-phonon frequencies\cite{Doniach1959,Kim1978,Varga1965,Mooradian1966}.
The effect is to cause a $q$ dependence near $\Gamma$ as:
\begin{equation}
  \omega^2_{\text{L},\nu}({q}) = \frac{q^2\omega^2_{\text{L},\nu} + q_{\infty,\text{scr}}^2\omega^2_{\text{T},\nu}}{q^2 + q_{\infty,\text{scr}}^2},
  \label{eq:loplasma}
\end{equation}
which goes to $\omega^2_{\text{T},\nu}$ at $q=0$, and approaches $\omega^2_{\text{L},\nu}$ for $q^2 \gg q_{\infty,\text{scr}}^2$.
We estimated the impact of including this $q$ dependence on the scattering rate for $10^{20}$ cm$^{-3}$ doping by calculating the average of the matrix element $|g_{{\bf q},\nu}|^2$ over $q$, and found it to make a difference of less than 5\%.
Therefore, to avoid complications in evaluating the energy conservation (discussed in Sec.~\ref{sec:compimp}), we neglect the $q$ dependence of the frequencies, and use the unscreened $q$-independent LO frequency values.

\subsection{Ionized impurity scattering} \label{sec:ionizedmethod}

To treat charged impurity scattering, two approaches are commonly used : Brooks-Herring (BH) or Conwell-Weisskopf (CW)\cite{Chattopadhyay1981}.
Both approaches are based on the Born approximation (elastic scattering), and differ only in the manner in which screening is treated.
The choice therefore depends on the screening regime, which is determined by the dimensionless parameter $\eta$:
\begin{equation}
  \eta = \frac{16 Z^2 N_\text{imp}^{2/3}}{q_{\text{scr}}^2} \frac{R_\text{H}^*}{\varepsilon_k} \, ,
\label{eq:eta}
\end{equation}
introduced by Ridley\cite{Ridley2001}, who suggested BH to be valid for $\eta < 1$, and CW for $\eta > 1$.
$R_\text{H}^*=13.605m_{\bf k}^*/m_e\epsilon^2$ eV is the effective Rydberg energy, where $m_{\bf k}^*$ is the band mass at wavevector ${\bf k}$ [discussed in Sec.~\ref{sec:compimp}].
The screening wavevector $q_{\text{scr}}^2$ is computed in a medium described by the static dielectric constant\cite{Chattopadhyay1981} $\epsilon$, taken to be 20 from Ref.~\onlinecite{stanislavchuk2012}, and is different from $q_{\infty,\text{scr}}^2$ used in the screening of electron-phonon interaction by a factor of $\epsilon_\infty/\epsilon$.
Based on Ridley's criterion, for the densities reported in this work ($10^{17}$--$10^{21}$ cm$^{-3}$) the BH approach should be applicable ($\eta < 1$).
The scattering rate for $N_\text{imp}$ ionized impurities of charge $Ze$ within the BH approach is given by
\begin{align*}
    &\tau_\text{imp}^{-1}({\bf k}) =
      \frac{\pi}{2} {v}_{\bf k} N_\text{imp} \left(\frac{Z e^2}{4\pi\epsilon\epsilon_0 \varepsilon_k} \right)^2 \\
      &\times \left[ \ln \left( 1 + \frac{8 m_{\bf k}^* \varepsilon_k}{\hbar^2 q_{\text{scr}}^2} \right)
      - \left( 1 +  \frac{\hbar^2 q_{\text{scr}}^2}{8 m_{\bf k}^* \varepsilon_k} \right)^{-1} \right]. \addtocounter{equation}{1}\tag{\theequation} \label{eq:imprate}
\end{align*}
BSO is often doped with substitutional La ($\text{La}_\text{Ba}^{+1}$) which has a $+1$ charge state\cite{Kim2013a}.
To simulate this situation we therefore assume ionized dopants to have $+1$ electronic charge, and their concentrations were chosen to be equal to the electron concentration, $N_\text{imp} = n$ (i.e., we assume full ionization, and no charge compensation).

For $Z > 1$, we would need fewer impurities to give rise to a given electron concentration $n$ (since $n = Z N_\text{imp}$, assuming complete ionization), but
because the scattering rate is proportional to $N_\text{imp}Z^2$ [Eq.~(\ref{eq:imprate})], the rate effectively increases linearly with $Z$.
Singly charged impurities are therefore optimal in terms of mobility.
For instance, doping with a double donor
would reduce the impurity-related part of the mobility by a factor of 2
compared to a single donor.

\subsection{Computational implementation} \label{sec:compimp}

The computation of mobility via Eq.~(\ref{eq:mu}) involves evaluating a three-dimensional (3-D) integral over the BZ.
The three quantities required for evaluating this integral are $v_{n,x}$, ${\partial f_{{\bf k}}}/{\partial \varepsilon_{{\bf k}}}$ and $\tau_{n}({\bf k})$.
The first two quantities converge reasonably well for finer grids, although working with such finer grids using hybrid functionals results in a prohibitively large computational cost unless an interpolation technique such as Wannier interpolation\cite{Marzari2012} is used.
However, the main bottleneck in computing the LO-phonon scattering process lies in obtaining $\tau_{n}({\bf k})$ [Eq.~(\ref{eq:tauinv})], which involves a 3-D integral over the phonon wavevector ${\bf q}$ with its integrand containing a $\delta$ function.
This integral can in principle be numerically evaluated from the full first-principles band structure.
In practice, however, any numerical technique employed will require a fine grid and the use of smearing to implement the energy-conserving $\delta$ function present in Eq.~(\ref{eq:tauinv}).
This leads to inaccuracies in the results because of its sensitivity to the choice of the smearing parameter\cite{Li2015,Yates2007}.
In addition, quantifying the error in the mobility is difficult without the knowledge of the exact result.

We circumvent these problems here by using an analytic expression for the CB dispersion.
This allows us to solve for one of the components of ${\bf q}$ using the condition for energy conservation {\it exactly},
thus reducing the integral to 2-D by getting rid of the $\delta$ function.
This has the added advantage of decreasing the computational complexity as well as producing the exact result that can be used to validate the choice of the smearing parameter in the numerical approach\cite{Yates2007,Li2015}.
Of course, this approach is contingent on the analytic expression being able to accurately reproduce the first-principles band structure, at least in the vicinity of the Fermi level, i.e., in the regions of the BZ where the factor ${\partial f_{{\bf k}}}/{\partial \varepsilon_{{\bf k}}}$ is non negligible.
The resulting integral can then be evaluated using any numerical integration technique; here we use the trapezoidal rule on a uniform grid separated by $\Delta k = 5\times10^{-3}$ \AA$^{-1}$ along each dimension.

\begin{figure}[!hc]
  \includegraphics[width=0.5\textwidth]{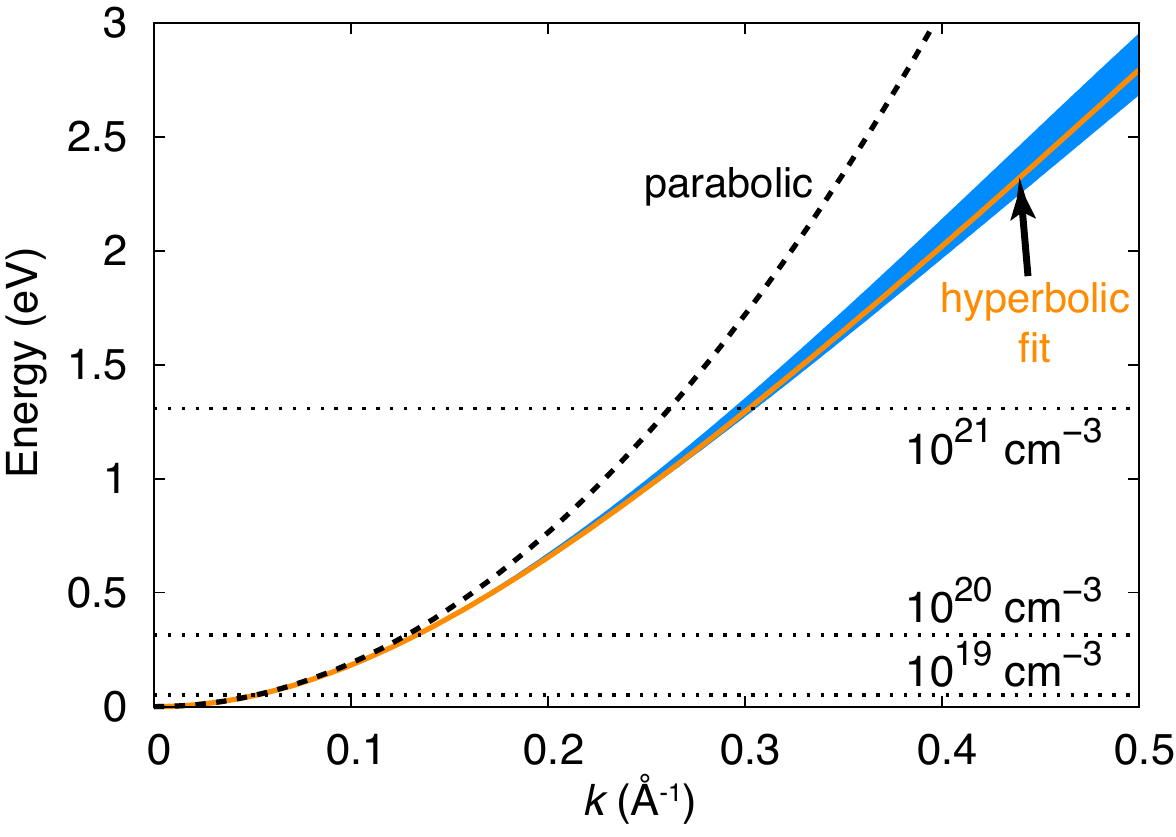}
  \caption{(Color online) Dispersion of the lowest CB around $\Gamma$, showing the slight anisotropy evident from the difference in dispersion along the high-symmetry directions ($\Gamma$-X, $\Gamma$-M, and $\Gamma$-R) (shaded in blue). The fitted hyperbolic dispersion (solid orange line) and parabolic dispersion (dashed black line) are shown.  Fermi levels for different electron densities are also indicated.}
  \label{fig:anisotropy}
\end{figure}

We plot the CB dispersion in Fig.~\ref{fig:anisotropy}.  It is clear that for energies larger than 0.3 eV nonparabolicity is significant.
Since the Fermi level may lie well above this energy for commonly used doping levels (also indicated in Fig.~\ref{fig:anisotropy})
using a parabolic dispersion relation would be inaccurate.
Instead, we use the hyperbolic dispersion relation derived from ${\bf k}.{\bf p}$ theory\cite{Dresselhaus2008}:
\begin{equation}
{\hbar^2k^2}/{2m_\Gamma^*} = \varepsilon_{{\bf k}}(1 + \alpha \varepsilon_{{\bf k}}) \, .
\label{eq:hyperbolic}
\end{equation}
Fitting our first-principles band structure of the CB to the hyperbolic dispersion relation
yields an effective mass near $\Gamma$, $m_\Gamma^* = 0.20\,m_e$, which we find to be isotropic.
However, the nonparabolicity parameter $\alpha$ was found to be slightly anisotropic as evident from inspecting the $E$-vs.-$k$ relation along the three high-symmetry directions in the BZ up to $k=0.4$ \AA$^{-1}$ (along $\Gamma\rightarrow$X, $\Gamma\rightarrow$M, and $\Gamma\rightarrow$R; see Table~\ref{table:mass}).
Since the deviation in $\alpha$ due to the anisotropy is small ($\pm0.03$ eV$^{-1}$), to avoid complications in our transport analysis, we use an isotropic $\alpha$ value of 0.25 eV$^{-1}$ determined by performing a weighted average of $\alpha$ along the high-symmetry directions.
Using this weighted-averaged value for $\alpha$ (weights of 6 along $\Gamma\rightarrow$X, 12 for M direction, and 8 for R direction) yields an accuracy of better than $\pm0.05$ eV for energies up to 1.5 eV ($k\sim0.32$ \AA$^{-1}$) above the CB minimum (see Fig.~\ref{fig:anisotropy}).
For reference, the Fermi level $\varepsilon_\text{F}$ corresponding to an electron concentration of 10$^{21}$ cm$^{-3}$ is at 1.34 eV; our fit will therefore be entirely adequate for all achievable doping densities.

\begin{table}[!h]
  \caption{The effective mass near $\Gamma$, $m_\Gamma^*$ and the nonparabolicity parameter $\alpha$ for the hyperbolic fit [Eq.~(\ref{eq:hyperbolic})] along three high-symmetry directions.}
\begin{ruledtabular}
  \begin{tabular}{ccc}
    \rule{0pt}{-3ex}Direction & $m_\Gamma^*$ ($m_e$) & $\alpha$ ($e$V$^{-1}$) \rule[-1.2ex]{0pt}{0pt} \\ \hline
  $\Gamma\rightarrow$X & 0.20 & 0.208 \\
  $\Gamma\rightarrow$M & 0.20 & 0.253 \\
  $\Gamma\rightarrow$R & 0.20 & 0.285 \\
\end{tabular}
\end{ruledtabular}
\label{table:mass}
\end{table}

For the hyperbolic dispersion, the band velocity is given by
\begin{align}
  {v}_{\bf k} = \frac{1}{\hbar}\frac{\partial \varepsilon_{\bf k}}{\partial {\bf k}} = \frac{\hbar {\bf k}}{m_\Gamma^*}\left( 1 + 2\alpha \varepsilon_{\bf k} \right)^{-1}
  \label{eq:velocity}
\end{align}
and the band mass accounting for nonparabolicity is given by
\begin{equation}
m_{\bf k}^* = m^*_\Gamma(1 + 2 \alpha\varepsilon_{\bf k})^3 \, .
\label{eq:mass}
\end{equation}
Our value for $m_\Gamma^*$ is in agreement with other HSE06 calculations\cite{scanlon-bso,Liu2013} except for one study\cite{Kim2013b}, in which the use of incorrect units for $k$ led to an overestimation of the mass (0.46$\,m_e$) by a factor of $(2\pi/a)^2$, where $a=4.13$ is the lattice parameter in \AA.
The value reported based on the hybrid functional PBE0\cite{scanlon-bso,Schlom-Piper2016} is 0.22$\,m_e$, and values based on LDA and GGA functionals\cite{Kim2013a,Moreira2012} range between 0.05 and 0.40$\,m_e$.
To our knowledge, values for the nonparabolicity parameter have not been reported.

To date, three studies have reported a wide range of experimental effective masses for the CB in BSO\cite{Kim2013a,Seo2014a,Allen2016}.
Kim {\it et al.}\cite{Kim2013a} estimated the mass to be 0.60$\,m_e$ based on the Burstein-Moss shift at $n=2.3\times10^{20}$ cm$^{-3}$.
Seo {\it et al.}\cite{Seo2014a} used the plasma frequency that fitted the observed Drude conductivity at an electron density of $8.9\times10^{19}$ cm$^{-3}$ to arrive at 0.35$\,m_e$.
Both these reports significantly overestimate $m_\Gamma^*$ compared to the calculated value of 0.20$\,m_e$.
However, the effective mass value of $\sim$0.20$\,m_e$ determined from reflectivity measurements by Allen {\it et al.}\cite{Allen2016} is in good agreement with our calculation.

As noted in Sec.~\ref{sec:LOphononmethod}, the screening wavevector is computed using Thomas-Fermi theory for a medium with a high-frequency dielectric constant $\epsilon_\infty$:
\begin{equation}
    q_{\infty,\text{scr}}^2 = \frac{e^2}{\epsilon_\infty\epsilon_0}\frac{\partial n}{\partial \varepsilon_\text{F}}=  \frac{e^2}{\epsilon_\infty\epsilon_0}\int_{\varepsilon_\text{CBM}}^\infty d\varepsilon \left(-\frac{\partial f}{\partial \varepsilon}\right) D\left(\varepsilon\right),
  \label{eq:qscr}
\end{equation}
where the density of states $D\left( \varepsilon \right)$ for the hyperbolic dispersion relation is given by
\begin{equation}
  D(\varepsilon) = \frac{1}{\pi^2} \left(\frac{m_\Gamma^*}{\hbar^2 }\right)^{3/2} \left( 1 + 2\alpha~\varepsilon \right)\sqrt{2\varepsilon\left( 1 + \alpha~\varepsilon\right)} .
  \label{eq:DOSvsE}
\end{equation}
The screening wavevector used in the calculation of ionized impurity scattering in Sec.~\ref{sec:ionizedmethod} is obtained as $q_{\text{scr}}^2 = \epsilon_\infty q_{\infty,\text{scr}}^2/\epsilon$.
The Fermi wavevector $k_\text{F}$ used in the Lindhard function $F(q/2k_\text{F})$ in Eq.~(\ref{eq:tauinv}) can be obtained by solving for $k$ in Eq.~(\ref{eq:hyperbolic}) at the Fermi energy $\varepsilon_\text{F}$.
For nondegenerate doping densities, where $\varepsilon_\text{F}$ lies below the CBM, we use the average energy of a classical gas $3k_\text{B}T/2$ to solve for $k_\text{F}$ using Eq.~(\ref{eq:hyperbolic}).

\subsection{Hall mobility and Hall factor}
The Hall mobility $\mu_\text{H}$ differs from the drift mobility $\mu$ by the Hall factor $r_\text{H}$ \cite{ziman1960book,Seegerbook}:
\begin{equation}
  \mu_\text{H} = r_\text{H}\mu \, .
  \label{eq:muH}
\end{equation}
The Hall factor, usually on the order of 1--2, is obtained from the electrical conductivity $\sigma$ [Eq.~(\ref{eq:mu})] and the conductivity coefficient $\sigma_\text{H}$ as\cite{Fu1996,Yates2007}
\begin{equation}
  r_\text{H} = \frac{\sigma_\text{H}}{\sigma^2} n e \, .
  \label{eq:rH}
\end{equation}
The conductivity coefficient is calculated using the expression
\begin{equation}
  \sigma_\text{H} = 2 e^3 \int_{\Omega} \frac{d^3k}{\left( 2\pi\right)^3} \tau^2\left({\bf k}\right) \left( - \frac{\partial f_{\bf k}}{\partial \varepsilon} \right) v_x\left( v_x M^{-1}_{yy} - v_y M^{-1}_{xy}\right),
  \label{eq:sigH}
\end{equation}
where, for the case of a hyperbolic band dispersion,
\begin{align}
  M^{-1}_{ij} &= \frac{1}{\hbar}\frac{\partial v_j}{\partial k_i} \\
  &= \left[m^*_\Gamma\left( 1 + 2\alpha\varepsilon_{\bf k} \right)\right]^{-1} \left( \delta_{ij} - 2\alpha m^*_\Gamma v_i v_j \right) \, .
  \label{eq:massij}
\end{align}
For parabolic dispersion, $\alpha =0$, $M^{-1}_{yy} = (m^*_\Gamma)^{-1}$ and $M^{-1}_{xy} = 0$.
Due to the dependence of $\sigma_\text{H}$ on $\tau^2({\bf k})$, $\sigma$ on $\tau({\bf k})$ and $r_\text{H}$ on their ratio, 
inclusion of the $k$-dependence of $\tau({\bf k})$ is essential for the calculation of Hall mobility.
Neglecting the $k$ dependence results in $r_\text{H} = 1$, i.e., the Hall and drift mobilities become identical.
\section{Results}
\label{sec:results}

Using the methodology described in the previous section, we now proceed to calculate the scattering rates and mobilities due to LO-phonon and ionized impurity scattering.
We first discuss the results obtained for the individual scattering processes separately.
Then, in Sec.~\ref{sec:totalmob}, we discuss the overall mobility combining the effect of both processes via Matthiessen's rule.

\subsection{LO-phonon scattering} \label{sec:LOphononresults}

\begin{figure}[!hc]
  \includegraphics[width=0.5\textwidth]{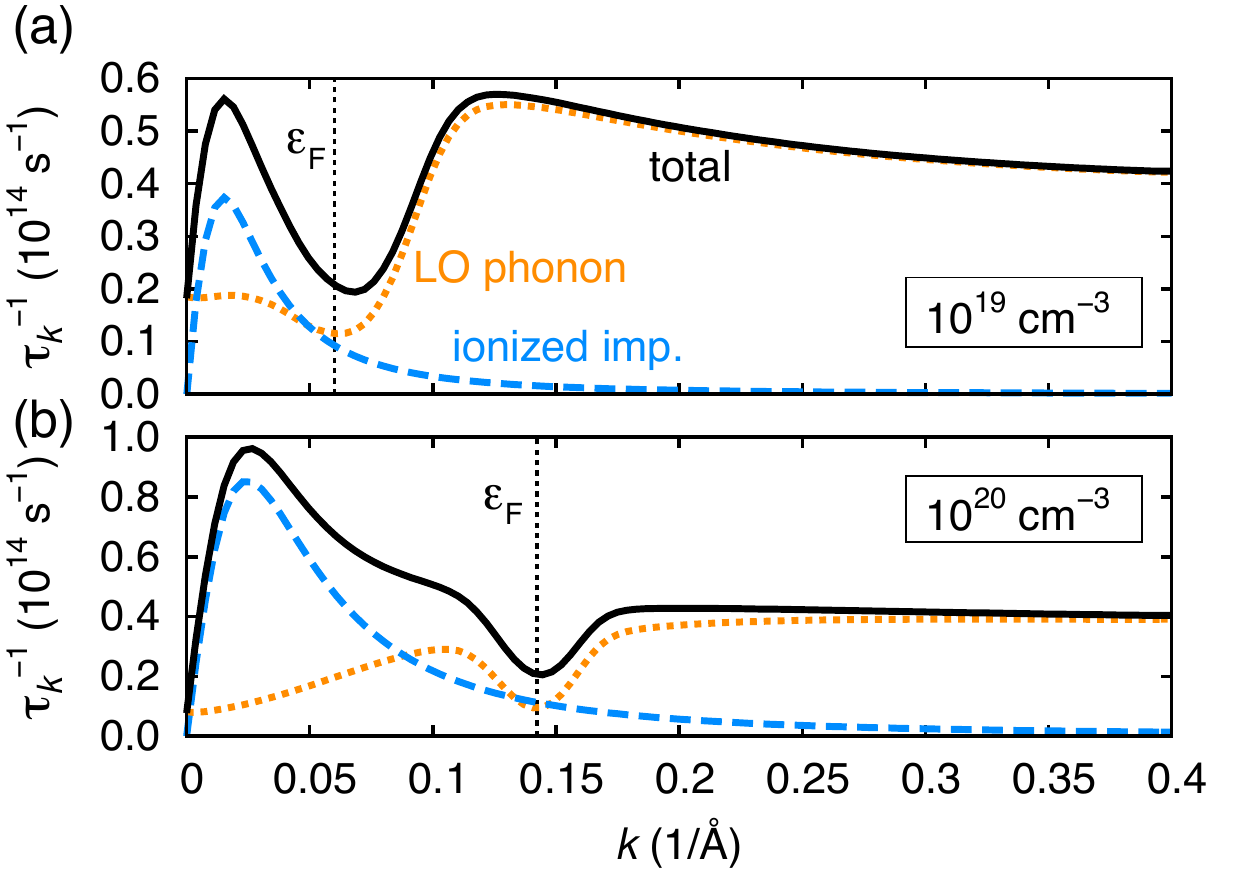}
  \caption{(Color online) Calculated scattering rates [10$^{14}$ s$^{-1}$] {\it versus} electron wavevector $k$ (\AA$^{-1}$) for LO-phonon scattering (orange dotted lines), ionized impurity scattering (blue dashed lines) and the total rate obtained via Matthiessen's rule (black solid lines) at RT (300 K) for electron densities (a) $10^{19}$ cm$^{-3}$ and (b) $10^{20}$ cm$^{-3}$.  Values are plotted along $\Gamma\rightarrow X$ but are representative of all directions in the BZ due to the almost isotropic band structure. Note the different vertical scales in panels (a) and (b).  The Fermi level $\varepsilon_\text{F}$ is indicated by vertical dashed lines.}
  \label{fig:tauX}
\end{figure}

The LO-phonon scattering rates are obtained using Eq.~(\ref{eq:tauinv}) by summing the contributions from the three polar LO modes.
The ${\bf k}$ dependence of the calculated scattering rate is plotted in Fig.~\ref{fig:tauX} for two values of the electron density, $10^{19}$ and $10^{20}$ cm$^{-3}$.
Our hyperbolic dispersion fit is valid over the plotted range (up to 0.4 \AA$^{-1}$), and since isotropic dispersion is a good approximation, the rate plotted is representative of all directions in the BZ.
We can explain the overall features in Fig.~\ref{fig:tauX} in terms of some basic mechanisms.
As $k$ increases, the band curvature decreases (see Fig.~\ref{fig:anisotropy}), and the radius of the energy surface at $\varepsilon_{\bf k + q}$ increases.
Both these characteristics cause the phonon wavevector ${\bf q}$ to be {\em larger} in order to satisfy the energy conservation, $\varepsilon_{\bf k + q} = \varepsilon_{\bf k} \pm \hbar\omega_\text{LO}$ due to the following two reasons:
(1) a smaller band curvature leads to a larger ${\bf k + q}$ for a given $\varepsilon_{\bf k + q}$, and
(2) a larger radius of the energy surface $\varepsilon_{\bf k + q}$ leads to an increase in the magnitude of $q$ for the {\em majority} of $q$ vectors satisfying energy conversation.
The ${\bf k}$ dependence of the scattering rate can therefore be related to the $q$ dependence in Eq.~(\ref{eq:epmatrix}).
We thus expect that near $k$=0, for $q$ values small compared to $q_{\infty,\text{scr}}$, the rate will be proportional to $q^2/q^4_{\infty,\text{scr}}$, while for larger $k$ values, if $q$ becomes larger than $q_{\infty,\text{scr}}$, the rate should decrease as $1/q^2$.
The behavior in Fig.~\ref{fig:tauX} is more complicated, however, due to the following reasons.
In Fig.~\ref{fig:tauX}(a), for $10^{19}$ cm$^{-3}$, the decrease at large $k$ can be observed beyond $k=0.15$ \AA$^{-1}$, but the initial rise near $k$=0 is overshadowed by the presence of a ``dip'' in the curve near the Fermi level; the origin of this feature will be explained below.
Figure~\ref{fig:tauX}(b), for $10^{20}$ cm$^{-3}$, does show the expected rise in the scattering rate for small $k$ values, but a decrease at large $k$ values is not evident. This is due to the large value of $q_{\infty,\text{scr}}$ in this case, which requires a much larger $q$, and hence a larger $k$, to observe the $1/q^2$ behavior.

The dip in the scattering rate around the Fermi level is a consequence of the energy dependence of the factor in square brackets in Eq.~(\ref{eq:tauinv}) that determines the probability of scattering due to emission or absorption of a phonon.
For a given phonon mode $\nu$ and Fermi level $\varepsilon_\text{F}$, $n_{{\bf q}\nu}$ is a constant (since phonon energy is independent of ${\bf q}$ in the Fr\"ohlich model), and therefore the ${\bf k}$ dependence of this factor is due only to the electronic occupation functions $f_{{\bf k}+{\bf q}}$ multiplied by their respective $\delta$ function for phonon absorption and emission.
With this information, this factor can be expressed solely in terms of the magnitude of the energy separation between the initial scattering state and the Fermi level, $|\varepsilon_{{\bf k}} - \varepsilon_\text{F}|$.
For initial states located exactly at the Fermi level ($\varepsilon_{{\bf k}} = \varepsilon_\text{F}$), this factor reaches a minimum at which the combined scattering probability due to emission and absorption is lowest, causing the dip in the scattering rate centered at the Fermi level.
For initial states located below the Fermi level ($\varepsilon_{{\bf k}} < \varepsilon_\text{F}$), the absorption term, which is proportional to the occupation of the final state dominates [see first term in Eq.~(\ref{eq:tauinv})].
The opposite is true for $\varepsilon_{{\bf k}} > \varepsilon_\text{F}$: the emission term becomes dominant when the final states are unoccupied due to the (1--$f_{{\bf k}+{\bf q}}$) term, which occurs for initial states occurring above $\varepsilon_\text{F}$.
As the doping is varied, the center of the dip in $\tau^{-1}({\bf k})$ moves along with the Fermi level, as is evident by comparing the ${\bf k}$ dependence of the rates for two different densities in Figs.~\ref{fig:tauX}(a) and (b).

Finally, we discuss the absolute value of the scattering rate at $\Gamma$.  This depends on the electron density because of
(1) screening, with the rate being proportional to $q^2/q^4_{\infty,\text{scr}}$ for small wavevectors, and
(2) the dip that moves along with the Fermi level, as discussed above.
For $n$=$10^{19}$ cm$^{-3}$ [Fig.~\ref{fig:tauX}(a)] the rate at $\Gamma$ is depressed partly due to the dip near $\varepsilon_\text{F}$, and partly due to screening.
At electron densities $10^{20}$ cm$^{-3}$ [Fig.~\ref{fig:tauX}(b)] and above, the Fermi level is pushed sufficiently high to ensure that the dip does not influence the value at $k$=0, and screening becomes the main effect near $\Gamma$; higher doping leads to more screening and hence decreases the rate [see Eq.~(\ref{eq:epmatrix})].

The calculated mobility values $\mu_\text{LO}$ based on the $k$-dependent LO-phonon scattering rates at RT (300 K) are listed in Table~\ref{table:mobility} for various doping levels.
Figure~\ref{fig:tauX} shows that the scattering rate in the vicinity of the Fermi level (which is what matters for mobility) can be different from the scattering rate at $\Gamma$.
Therefore, using a scattering rate calculated at $\Gamma$ in a constant-$\tau$ approximation could lead to inaccurate mobilities as well as incorrect trends with electron density (or $\varepsilon_\text{F}$).
For example, at RT for $n = 10^{20}$ cm$^{-3}$, the mobility calculated using a $k$-independent scattering rate with the value evaluated at $\Gamma$ ($\tau_\Gamma^{-1} = 0.080\times10^{14}$ s$^{-1}$) is 943 cm$^{2}$V$^{-1}$s$^{-1}$.
This value overestimates the actual mobility (calculated taking the ${\bf k}$ dependence into account), 594 cm$^2$V$^{-1}$s$^{-1}$.
Based on the factor ${\partial f_{{\bf k}}}/{\partial \varepsilon_{{\bf k}}}$ in Eq.~(\ref{eq:mu}), which renders the mobility sensitive only to rates in the vicinity of $\varepsilon_\text{F}$, one would expect a constant $\tau$ evaluated at $\varepsilon_\text{F}$ ($\tau_{k_\text{F}}$) to yield more accurate values.
However, we find that using $\tau_{k_\text{F}}^{-1}$ ($= 0.095\times10^{14}$ s$^{-1}$) yields 792 cm$^2$V$^{-1}$s$^{-1}$, which still significantly overestimates the actual mobility by about 33\% (200 cm$^2$V$^{-1}$s$^{-1}$).
On the other hand, for $n = 10^{19}$ cm$^{-3}$, we find a different trend: using $\tau_\Gamma^{-1}$ ($= 0.186\times10^{14}$ s$^{-1}$) gives a reasonable mobility of 455 cm$^{2}$V$^{-1}$s$^{-1}$ compared to the actual mobility of 487 cm$^2$V$^{-1}$s$^{-1}$, whereas using $\tau_{k_\text{F}}^{-1}$ ($= 0.116\times10^{14}$ s$^{-1}$) yields 733 cm$^2$V$^{-1}$s$^{-1}$, which is a severe overestimation by about 50\% (246 cm$^2$V$^{-1}$s$^{-1}$).

\begin{table}[!h]
  \caption{Calculated drift mobility values (in cm$^2$V$^{-1}$s$^{-1}$) at RT (300 K) taking into account scattering due to LO phonons ($\mu_\text{LO}$), ionized dopants ($\mu_\text{imp}$), and their total ($\mu_\text{tot}$) for different electron densities $n$ (cm$^{-3}$). The corresponding Fermi level $\varepsilon_\text{F}$ (eV) (referenced to the conduction-band minimum, CBM), Fermi wavevector $k_\text{F}$ (\AA$^{-1}$), and the screening wavevectors $q_{\infty,\text{scr}}$ (\AA$^{-1}$) [Eq.~(\ref{eq:qscr})] and $q_{\text{scr}}$ (\AA$^{-1}$) are also given.}
\begin{ruledtabular}
  \begin{tabular}{ccccccrc}
    \rule{0pt}{-3ex}n & $\varepsilon_\text{F} - \varepsilon_\text{CBM}$ & $k_{\text{F}}$ & $q_{\infty,\text{scr}}$ & $q_{\text{scr}}$ & $\mu_\text{LO}$ & $\mu_\text{imp}$ & $\mu_\text{tot}$\rule[-1.2ex]{0pt}{0pt} \\ \hline
  \rule{0pt}{3ex}
  10$^{17}$ & -0.079 & 0.036 & 0.013 & 0.006 & 321 & 29850 & 307 \\
  10$^{18}$ & -0.017 & 0.036 & 0.038 & 0.018 & 389 & 5195  & 318 \\
  10$^{19}$ & 0.074  & 0.063 & 0.083 & 0.039 & 487 & 1445  & 329 \\
  10$^{20}$ & 0.354  & 0.142 & 0.137 & 0.063 & 594 & 666   & 305 \\
  10$^{21}$ & 1.343  & 0.307 & 0.240 & 0.111 & 530 & 290   & 183 \\
\end{tabular}
\end{ruledtabular}
\label{table:mobility}
\end{table}

We note that for nondegenerate doping, where the Fermi level lies in the band gap, the reasoning based on ${\partial f_{{\bf k}}}/{\partial \varepsilon_{{\bf k}}}$ leads us to expect that using $\tau_\Gamma$ should give reasonably accurate mobility values.
Indeed, for $n$=$10^{17}$ cm$^{-3}$, using $\tau^{-1}_\Gamma = 0.263\times10^{14}$s$^{-1}$ gives a mobility of 334 cm$^{2}$V$^{-1}$s$^{-1}$, in good agreement with the ${\bf k}$-dependent $\tau$ calculation value 321 cm$^{2}$V$^{-1}$ s$^{-1}$.
However, for $n$=$10^{18}$ cm$^{-3}$, $\tau^{-1}_\Gamma = 0.178\times10^{14}$ s$^{-1}$ results in 487 cm$^{2}$V$^{-1}$s$^{-1}$, a serious overestimate compared to the actual value of 389 cm$^{2}$V$^{-1}$s$^{-1}$ obtained using the full ${\bf k}$ dependence of $\tau$.
This example demonstrates that using the rate at $\Gamma$ is a good approximation only for low doping concentrations, corresponding to Fermi levels well below the CBM, where only carriers very close to $\Gamma$ contribute to transport.
For higher doping levels there is no justification for using a constant-$\tau$ value determined either at the $\Gamma$ or near $\varepsilon_\text{F}$.

\begin{figure}[!hc]
  \includegraphics[width=0.5\textwidth]{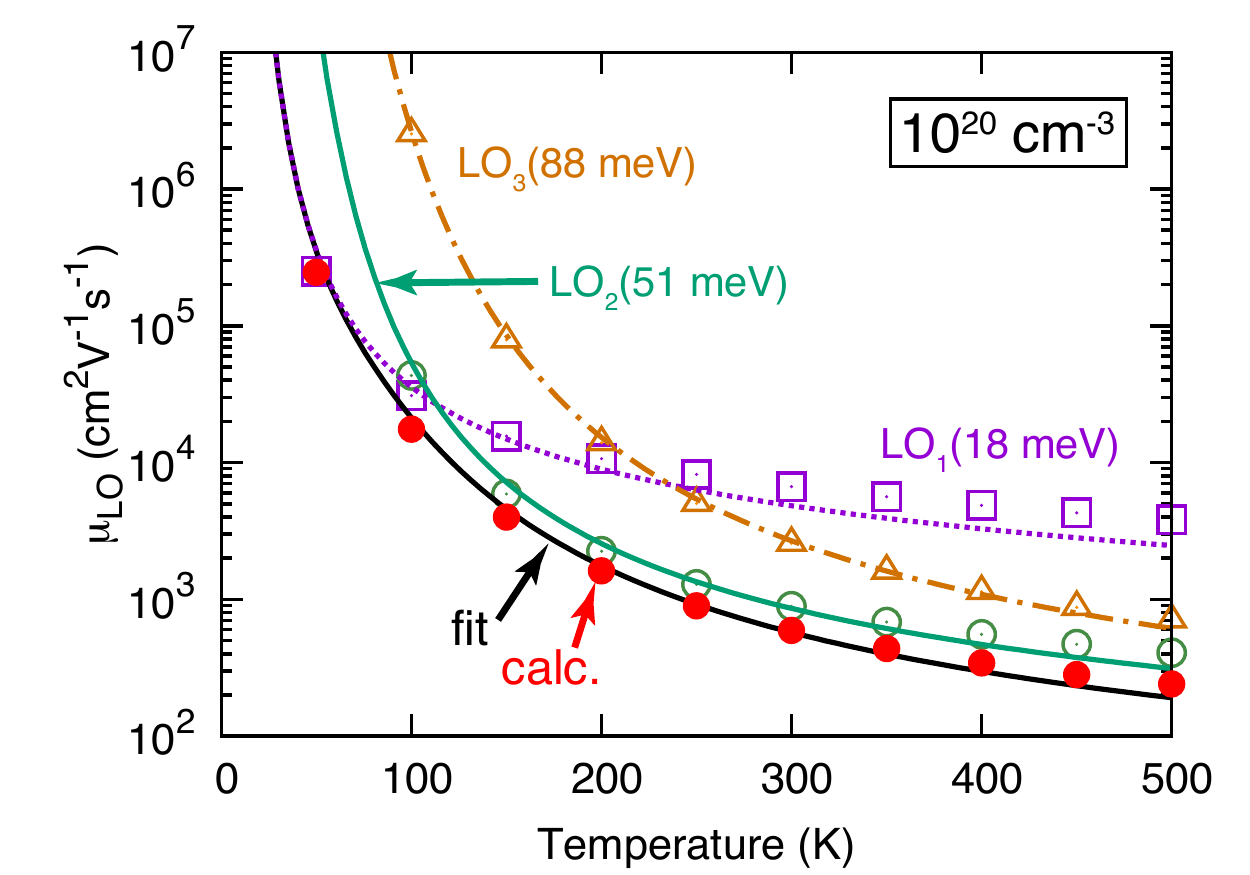}
  \caption{(Color online) Calculated mobility {\it versus} temperature for LO-phonon scattering ($\mu_\text{LO}$) (solid red circles) for $n$=$10^{20}$ cm$^{-3}$.  The calculated mobility due to scattering from the individual phonon modes is shown: $\mu_\text{LO1}$ (purple open squares), $\mu_\text{LO2}$ (green open circles), and $\mu_\text{LO3}$ (orange triangles) with energies 18, 51 and 88 m$e$V, respectively. The lines are analytic fits for the mobilities based on the BE distribution (see text): $\mu_\text{LO}$ (black solid line), $\mu_\text{LO1}$ (purple dotted lines), $\mu_\text{LO2}$ (green dashed line), and $\mu_\text{LO3}$ (orange dash-dot line).}
  \label{fig:mobvsT-LO}
\end{figure}

Figure~\ref{fig:mobvsT-LO} shows the temperature dependence of the mobility based on LO-phonon scattering: the mobility decreases by more than two orders of magnitude going from 50 K to 300 K.
This strong dependence results from the phonon occupation factor $n_{{\bf q}\nu}$ entering into the LO-phonon scattering rate [Eq.~(\ref{eq:tauinv})].
This behavior as a function of temperature is similar for all electron concentrations.
Figure~\ref{fig:mobvsT-LO} also shows the contributions to the mobility due to the individual LO phonon modes.
As expected, at low temperatures (0--100 K), only the lowest frequency mode (LO$_1$) is occupied and contributes to limiting the mobility.
Starting at 100 K the LO$_2$ mode (51 meV) gets populated.
The highest energy (88 m$e$V) LO$_3$ mode starts contributing to scattering at temperatures above 250 K.

LO-phonon contributions to mobility are often fitted\cite{Verma2014a,Mikheev2015,Frederikse1967} to an expression that is inversely proportional to the BE distribution as derived by Low and Pines\cite{Low1955} for a single LO mode.
For materials with multiple LO modes, such as the perovskite oxides, the fits are performed by adding the reciprocal mobilities due to each mode with some assumption or knowledge about which modes dominate in the temperature range of interest\cite{Frederikse1967,Verma2014a}.
To assess the validity of such a procedure, we fitted our calculated mobilities due to the individual LO phonon modes to a BE-distribution-like term, as well as their combined mobility by summing their reciprocals:
\begin{equation}
  \mu^{-1}_\text{LO} = \mu^{-1}_{\text{LO}_1} + \mu^{-1}_{\text{LO}_2} + \mu^{-1}_{\text{LO}_3} = \sum_\nu \frac{c_\nu}{\exp\left(\frac{\hbar\omega_{\nu}}{k_\text{B} T}\right) - 1},
  \label{eq:BEdist}
\end{equation}
where $c_\nu$ is the fitting coefficient corresponding to the phonon mode $\nu$ with energy $\hbar\omega_{\nu}$.
The fits for the individual modes (see Fig.~\ref{fig:mobvsT-LO}) perform very well when $k_\text{B}T \ll \hbar\omega_\text{LO}$, consistent with the fact that the expression was derived in the low-temperature limit by Low and Pines\cite{Low1955}.
At higher temperatures, the fits tend to slightly underestimate the mobility.

\subsection{Ionized impurity scattering} \label{sec:ionizedresults}

\begin{figure}[!hc]
  \includegraphics[width=0.5\textwidth]{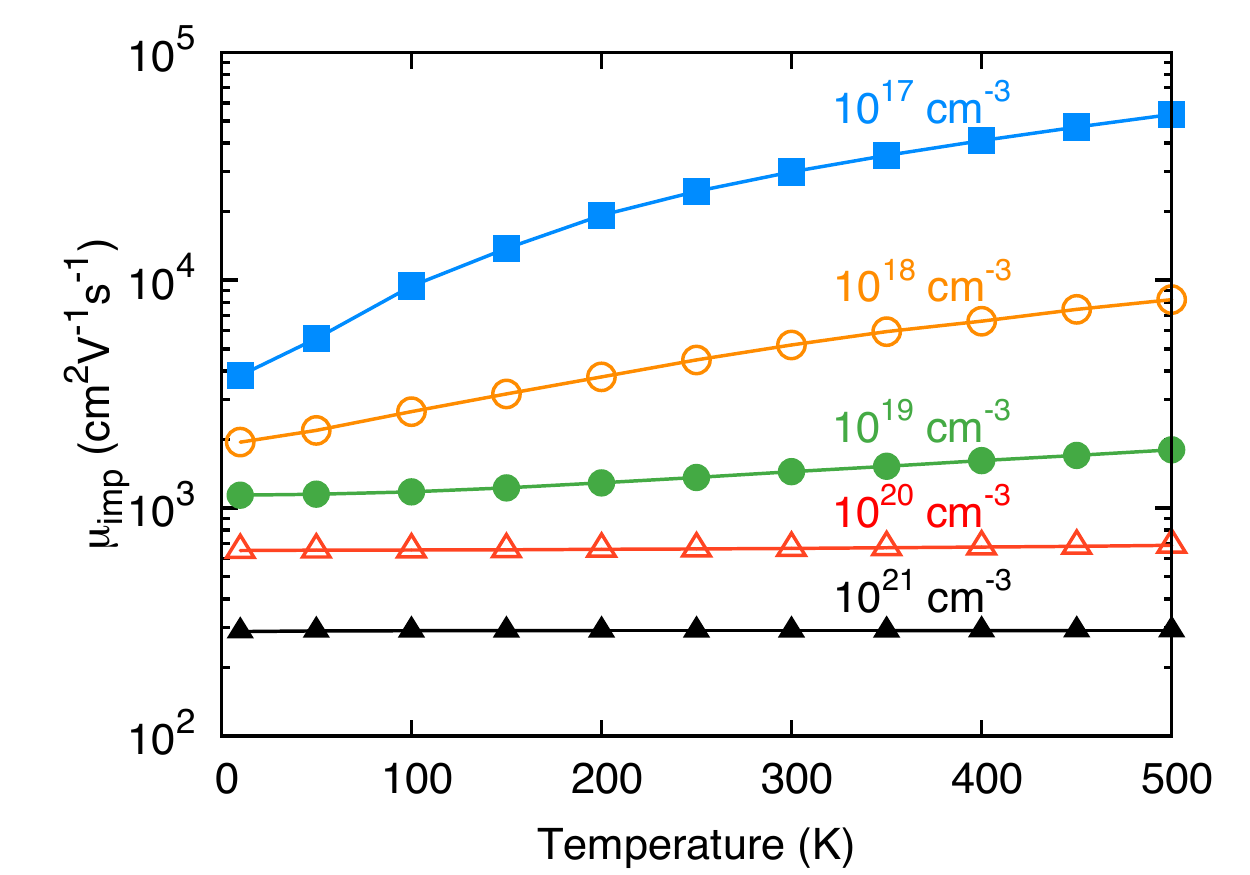}
  \caption{(Color online) Calculated drift mobility {\it versus} temperature in the case of ionized impurity scattering ($\mu_\text{imp}$) for five different electron densities: $10^{17}$ (blue squares), $10^{18}$ (orange open circles), $10^{19}$ (green solid circles), $10^{20}$ (red open triangles), and $10^{21}$ (black closed triangles) cm$^{-3}$.}
  \label{fig:mobvsT-imp}
\end{figure}

We now proceed to calculate ionized impurity scattering, based on Eq.~(\ref{eq:imprate}).
The ${\bf k}$ dependence of the scattering rate is shown in Fig.~\ref{fig:tauX}.  We observe that the behavior is linear in $k$ near $\Gamma$, and decreases as $k^{-3}$ beyond the peak.
The resulting mobility values calculated at RT are included in Table~\ref{table:mobility} for doping densities $10^{17}$--$10^{21}$ cm$^{-3}$.
Figure~\ref{fig:mobvsT-imp} shows the temperature dependence of the drift mobility for various doping densities.
For doping densities below $10^{19}$ cm$^{-3}$, the mobility depends strongly on temperature, while for higher doping densities the mobility is temperature independent.
Inspection of Eq.~(\ref{eq:mu}) shows that the temperature dependence arises from ${\partial f_{{\bf k}}}/{\partial \varepsilon_{{\bf k}}}$ centered around the Fermi level $\varepsilon_\text{F}$ with a temperature-dependent width.
The screening wavevector $q_{\text{scr}}$ appearing in Eq.~(\ref{eq:imprate}) is also dependent on ${\partial f_{{\bf k}}}/{\partial \varepsilon_{{\bf k}}}$ [see Eq.~(\ref{eq:qscr})].
Therefore, any temperature dependence in the mobility should come primarily from ${\partial f_{{\bf k}}}/{\partial \varepsilon_{{\bf k}}}$ and $\varepsilon_\text{F}$.

As the temperature decreases two changes occur: (1) the width of ${\partial f_{{\bf k}}}/{\partial \varepsilon_{{\bf k}}}$ narrows and approaches a $\delta$ function at 0 K, and (2) the Fermi level $\varepsilon_\text{F}$ itself increases.
Both these dependences affect the mobilities when the scattering rate varies rapidly with energy (or $k$) around the Fermi level.
This is indeed the case for the ionized impurity scattering rate close to $\Gamma$, i.e., for $\varepsilon_\text{F}$ near or below the CBM.
Test calculations for $10^{18}$ and $10^{19}$ cm$^{-3}$ in which the width of the Fermi derivative ${\partial f_{{\bf k}}}/{\partial \varepsilon_{{\bf k}}}$ as well as $\varepsilon_\text{F}$ itself were fixed to their values at 300 K confirmed this reasoning; the resulting mobilities showed no temperature dependence.
At doping densities above $10^{19}$ cm$^{-3}$ the variation of the rate around the Fermi level is slow, resulting in a very weak temperature dependence of the mobilities.

\subsection{Total drift mobility} \label{sec:totalmob}

\begin{figure}[!hc]
  \includegraphics[width=0.5\textwidth]{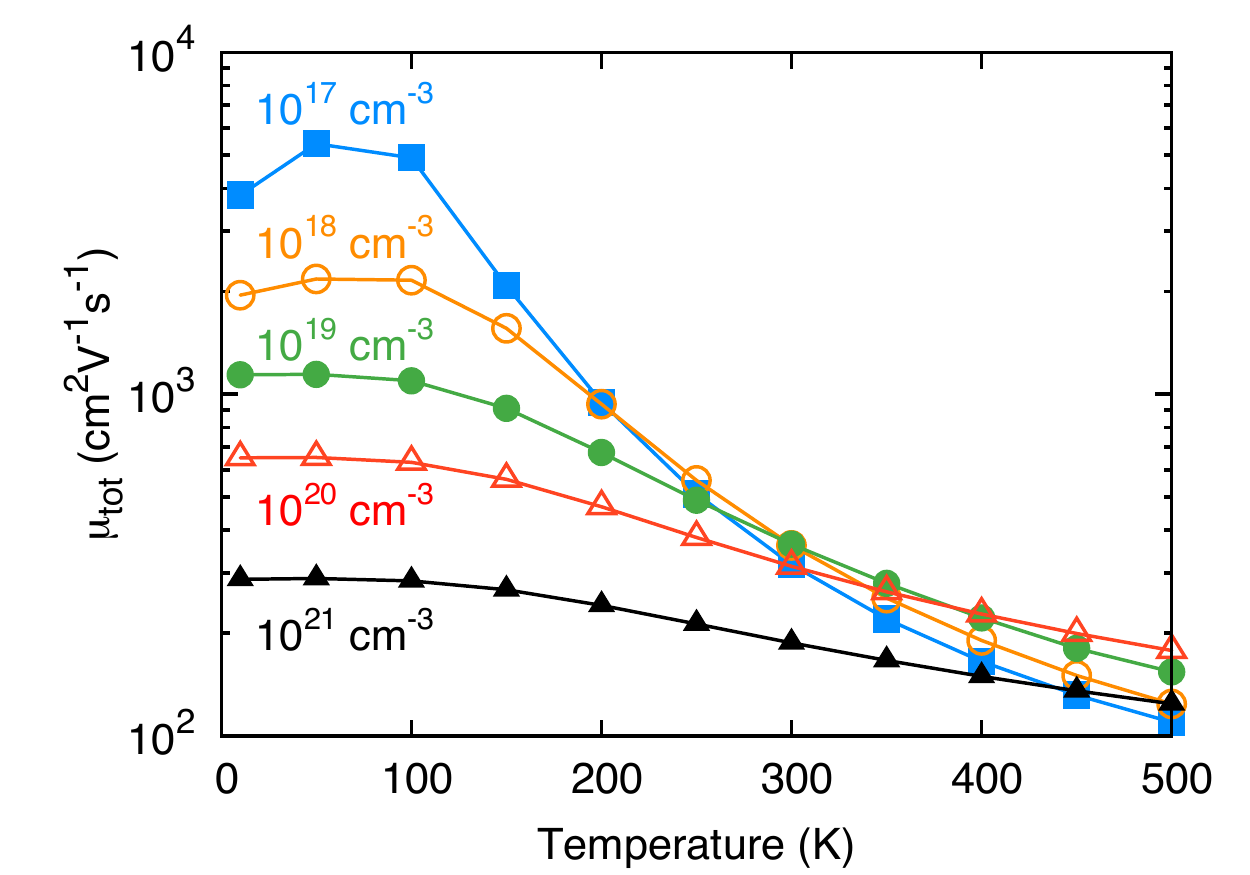}
  \caption{(Color online) Calculated drift mobility {\it versus} temperature due to a combination of LO-phonon and ionized impurity scattering ($\mu_\text{tot}$) for five different electron densities: $10^{17}$ (blue squares), $10^{18}$ (orange open circles), $10^{19}$ (green solid circles), $10^{20}$ (red open triangles), and $10^{21}$ (black closed triangles) cm$^{-3}$.}
  \label{fig:mobvsT-tot}
\end{figure}

We now combine the effect of LO-phonon and ionized impurity scattering via Matthiessen's rule, $\tau^{-1}_\text{tot} = \tau^{-1}_\text{LO} + \tau^{-1}_\text{imp}$.
The temperature dependence of the total drift mobility [see Fig.~\ref{fig:mobvsT-tot}] shows the typical behavior\cite{Yu2005}:
ionized impurity scattering dominates at low temperatures, and as the temperature increases LO-phonon scattering reduces the mobility.
At RT and for doping densities less than $10^{18}$ cm$^{-3}$, the total mobility (see Table~\ref{table:mobility}) is limited mainly by LO-phonon scattering.  At higher doping levels impurity scattering plays an increasingly important role. This highlights the importance of efforts to reduce ionized impurity scattering, as discussed in Sec.~\ref{sec:enhancemob}.

\subsection{Mobility {\it vs.} electron density} \label{sec:mob-vs-den}

\begin{figure}[!hc]
  \includegraphics[width=0.5\textwidth]{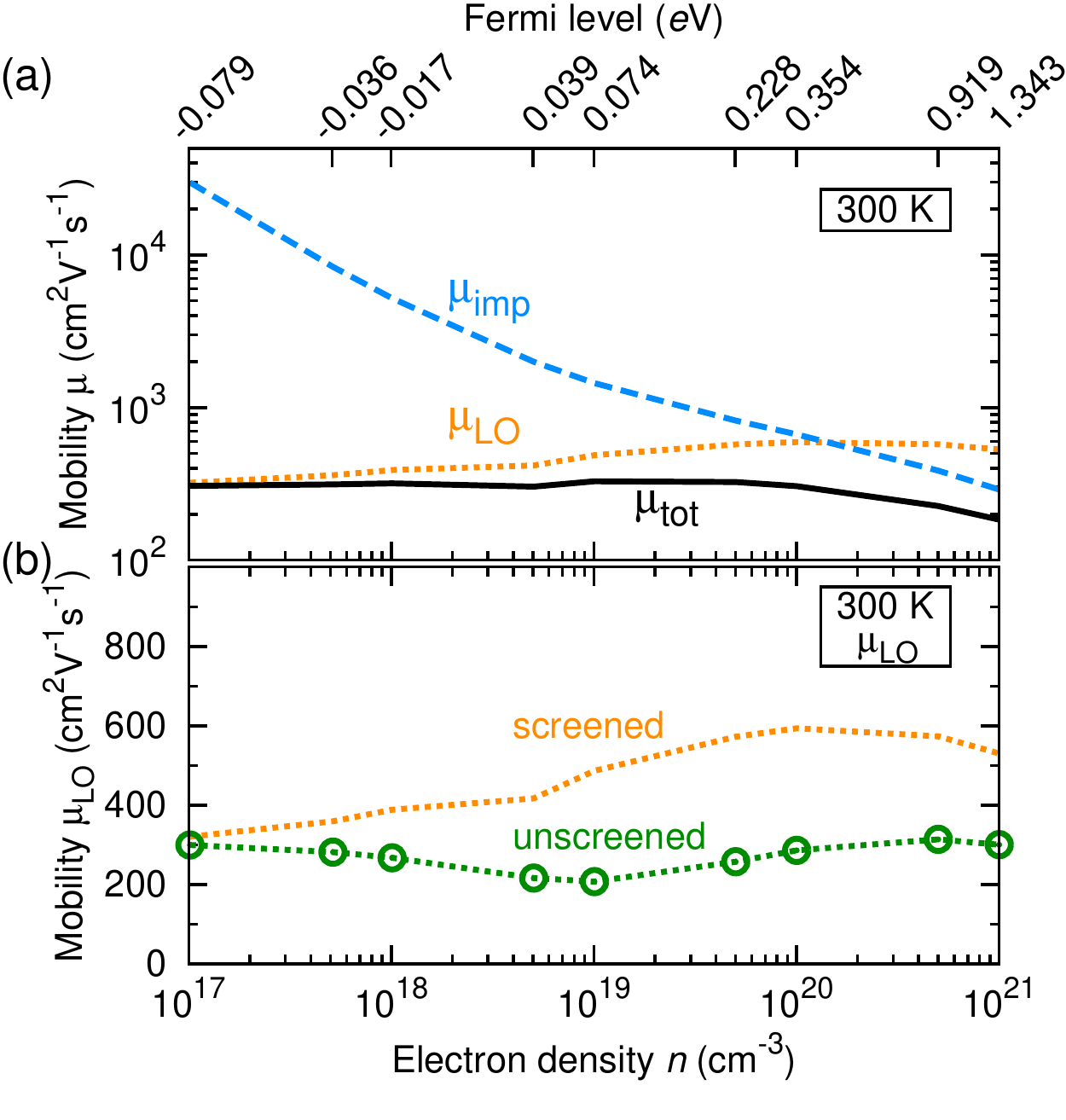}
  \caption{(Color online) (a) Calculated drift mobility {\it versus} electron density (cm$^{-3}$) at 300 K (RT) for LO-phonon scattering, $\mu_\text{LO}$ (orange dotted line), and ionized impurity scattering, $\mu_\text{imp}$ (blue dashed line), as well as the total drift mobility, $\mu_\text{tot}$ (black solid line). (b) Comparison of the screened (orange dotted line) and unscreened (green open circles on dotted line) values for $\mu_\text{LO}$ {\it versus} electron density.}
  \label{fig:mobvsden}
\end{figure}

The trends in the RT mobility-vs.-density ($\mu$-vs.-$n$) as well as the dominant scattering mechanism at a given doping level can be visualized more clearly from Fig.~\ref{fig:mobvsden}.
First, we discuss the trends for the two scattering mechanisms individually.
Ionized impurities are seen to reduce the mobility with doping as $n^{m}$ (linear on the log-log plot), where the exponent $m$ is different for the two identifiable density regimes: (1) nondegenerate doping: $n < 10^{18}$ cm$^{-3}$, when the Fermi level lies in the band gap, where the mobility decreases as $\sim$$n^{-0.75}$, and (2) degenerate doping: $n > 5\times10^{18}$ cm$^{-3}$, when the Fermi level is above the CBM, where the mobility decreases more slowly as $\sim$$n^{-0.33}$.
As discussed in Sec.~\ref{sec:ionizedmethod}, we have set $n=N_\text{imp}$ for an ionized donor of +1 charge.
In the nondegenerate doping regime, screening is weak ($q_{\text{scr}}$ is small) and the factor outside the square brackets in Eq.~(\ref{eq:imprate}) dominates and yields a $N_\text{imp}^{-1}$ dependence, close to but not quite equal to the $N_\text{imp}^{-0.75}$ behavior extracted from the full results in Fig.~\ref{fig:mobvsden}.
In the degenerate doping regime, screening is significant(large value of $q_{\text{scr}}$), and the factor within square brackets in Eq.~(\ref{eq:imprate}) becomes important.
An expansion of Eq.~(\ref{eq:imprate}) for large $q_{\text{scr}}$ shows that the mobility should decreases as $N_\text{imp}^{-1}q^4_{0,\text{scr}}\propto n^{-1/3}$, in agreement with the behavior in Fig.~\ref{fig:mobvsden}(a).
Here we have used $q_\text{scr}\sim n^{1/6}$ from Thomas-Fermi theory for degenerate doping.

For LO-phonon scattering, we find that screening plays a significant role for densities $10^{18}$ cm$^{-3}$ and higher, as seen by comparing mobilities with and without screening in Fig.~\ref{fig:mobvsden}(b).
To elucidate the role played by the band structure and the Fermi level, we focus on the {\it unscreened} case. Based on the physics, three distinct regions can be identified in the $\mu$-vs.-$n$ curve:
(1) for low densities $\le 10^{18}$ cm$^{-3}$, the mobility is fairly constant, at $\sim$300 cm$^2$V$^{-1}$s$^{-1}$;
(2) for densities in the range $10^{18}$--$10^{19}$ cm$^{-3}$, the mobility decreases slightly to $\sim$200 cm$^2$V$^{-1}$s$^{-1}$; and
(3) for doping densities $>10^{19}$ cm$^{-3}$, the mobility increases with electron density from $\sim$200 to $\sim$300 cm$^2$V$^{-1}$s$^{-1}$.
The effect of screening is to increase the mobility with increasing density.

In Region 1 the Fermi level lies below $\varepsilon_\text{CBM} + \hbar\omega_\text{LO1}$, where $\hbar\omega_\text{LO1}=18$ meV is the lowest LO-phonon energy among the three LO modes.
This makes emission of LO phonon almost impossible, since most of the carriers are at the CBM and do not have any states below to relax to after emission.
Phonon absorption remains as the only LO-phonon scattering process, and determines the mobility.
As the Fermi level moves above $\varepsilon_\text{CBM} + \hbar\omega_\text{LO1}$ with doping, we enter Region 2, where LO-phonon emission becomes possible and reduces the mobility.
Raising the Fermi level further (Region 3) results in an {\it increase} in mobility, which can be attributed to a decrease in band curvature and a larger energy surface as discussed in Sec.~\ref{sec:LOphononresults}.
In the presence of screening, for densities greater than $10^{18}$ cm$^{-3}$, it is the screening wavevector $q_{\infty,\text{scr}}$ that mainly determines the mobility by enhancing it.

Overall, however, the {\it total} mobility is seen to decrease with increasing electron density [Fig.~\ref{fig:mobvsden}(a)] due to the strong contribution from ionized impurity scattering.
On comparing $\mu_\text{LO}$ and $\mu_\text{imp}$, it is clear that ionized dopants affect the RT mobility already at densities $n > 10^{18}$ cm$^{-3}$, and are the dominant source of scattering for $n > 10^{20}$ cm$^{-3}$.
For $n > 10^{18}$ cm$^{-3}$, dopants in combination with LO-phonon scattering limit the total mobility to less than 330 cm$^2$V$^{-1}$s$^{-1}$.
Below $n = 10^{18}$ cm$^{-3}$, the limit is determined by LO-phonon scattering, but the lack of screening results in a lower mobility of $\sim$300 cm$^2$V$^{-1}$s$^{-1}$.
Of course, reducing the doping would also reduce the conductivity;
it is therefore important to consider doping techniques that can mitigate impurity scattering without sacrificing the carrier concentration and conductivity.
We will discuss two such techniques in Sec.~\ref{sec:enhancemob}.

\section{Discussion} \label{sec:discussion}

\subsection{Comparison with experimental measurements} \label{sec:compexp}
Our discussions thus far have focused on the drift mobility.
However, all of the transport measurements on BSO\cite{Kim2012,kim-bso-mobility-2,Kim2013b,Raghavan2016,Schlom-Piper2016}, with the exception of one report of transistor-based measurements\cite{Kim2015} on thin films, have been based on Hall measurements.
To allow for a direct quantitative comparison with experiments we need to compute the Hall mobility, which is related to the drift mobility $\mu$ via the Hall factor $r_\text{H}$ as given in Eq.~(\ref{eq:muH}).

\begin{figure}[!hc]
  \includegraphics[width=0.5\textwidth]{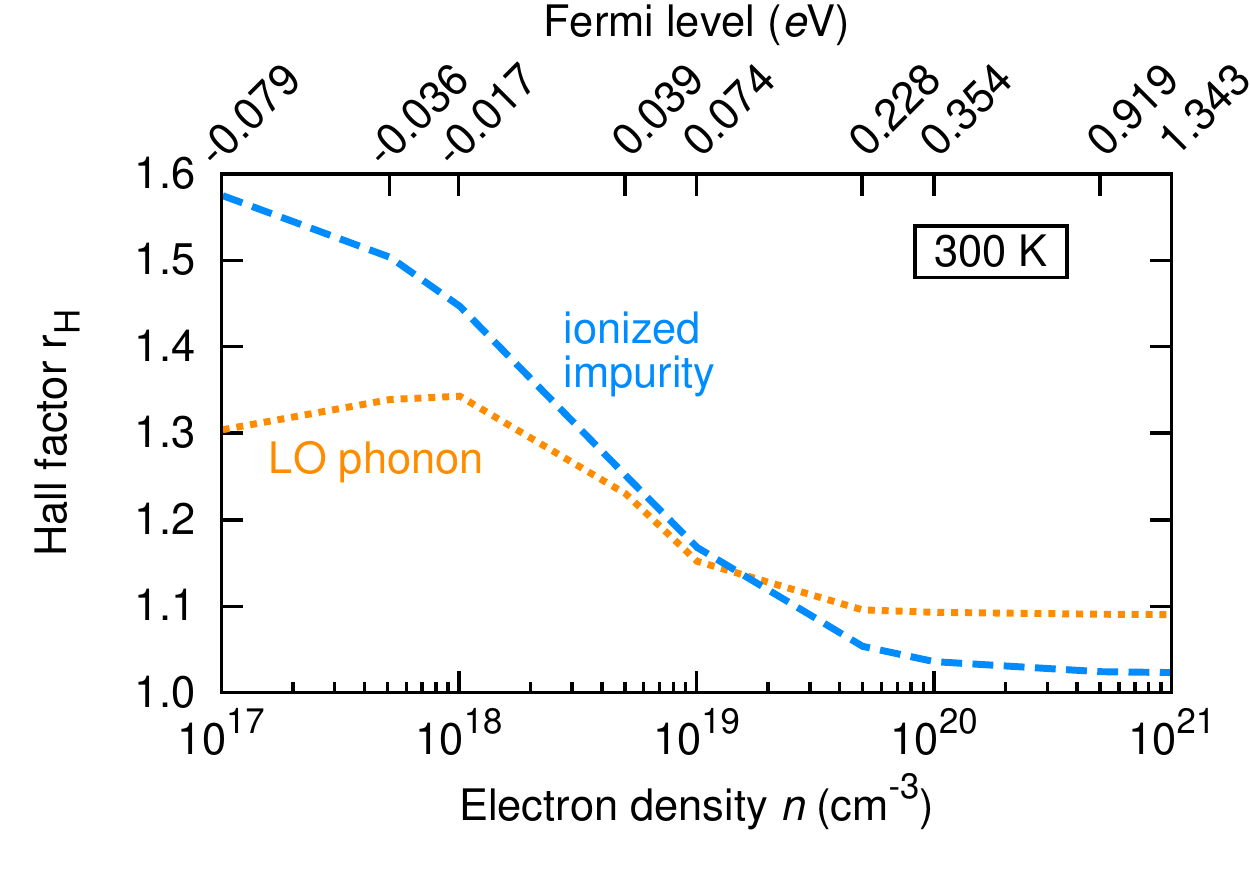}
  \caption{(Color online) Calculated Hall factor $r_\text{H}$ {\it versus} carrier concentration for ionized impurity (blue dashed line) and LO-phonon scattering (orange dotted line).}
  \label{fig:rH-300K}
\end{figure}

Due to its dependence on $\tau\left({\bf k}\right)$, the Hall factor $r_\text{H}$ calculated using Eq.~(\ref{eq:rH}) depends on the scattering process, and shows a strong dependence on carrier concentration (see Fig.~\ref{fig:rH-300K}) as well as temperature.
With increasing carrier concentration, as the Fermi level approaches the CBM and moves above into the CB, $r_\text{H}$ decreases and saturates to a constant value of $\sim1.09$ for LO-phonon scattering, and $\sim1.03$ for ionized impurity scattering.

We now explicitly compare the temperature dependence of our calculated Hall mobility $\mu_\text{H}$ with experiment.  We focus on the highest mobility values reported to date, from experiments on bulk single crystals\cite{kim-bso-mobility-2} as well as thin films\cite{Raghavan2016}.

\subsubsection{Bulk crystals}\label{sec:compexp-bulk}
\begin{figure}[!hc]
  \includegraphics[width=0.5\textwidth]{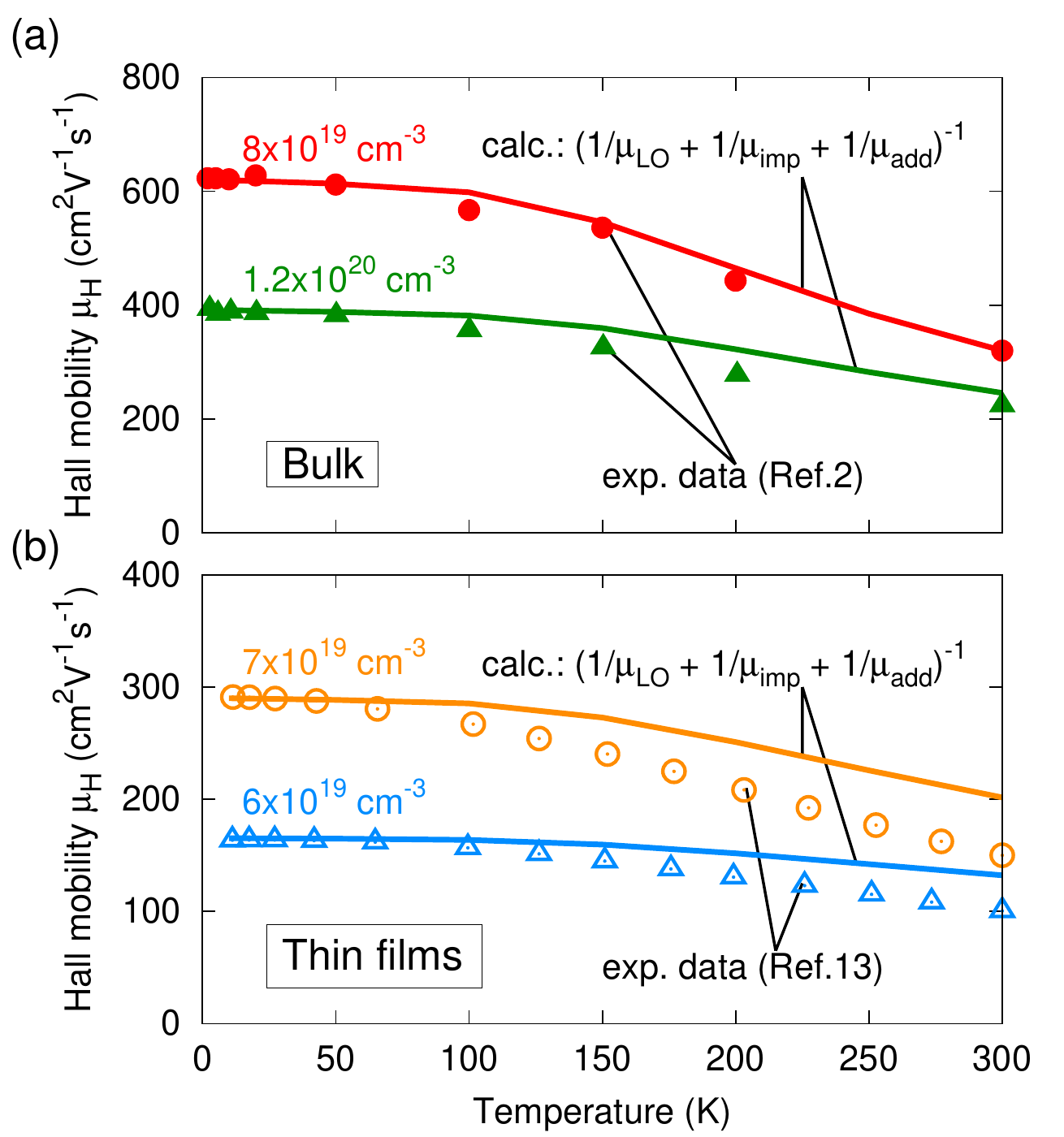}
  \caption{(Color online) Solid lines: calculated Hall mobility {\it versus} temperature due to ionized impurity and LO-phonon scattering, with addition of a temperature-independent scattering contribution: $(1/\mu_\text{imp} + 1/\mu_\text{LO} + 1/\mu_\text{add})^{-1}$, for different experimental doping densities.
  Symbols indicate the corresponding experimental Hall measurements (a) on bulk crystals from Ref.~\onlinecite{kim-bso-mobility-2} , for $8\times10^{19}$ cm$^{-3}$ (solid red circles) and $1.2\times10^{20}$ cm$^{-3}$ (solid green triangles) doping, and
  (b) on thin films from Ref.~\onlinecite{Raghavan2016}, for $6\times10^{19}$ cm$^{-3}$ (open orange circles) and $7\times10^{19}$ cm$^{-3}$ (open blue triangles) doping.
  For bulk crystals (a), $\mu_\text{add}$ is 4500 cm$^2$V$^{-1}$s$^{-1}$ for $8\times10^{19}$ cm$^{-3}$, and 1000 cm$^2$V$^{-1}$s$^{-1}$ for $1.2\times10^{20}$ cm$^{-3}$, and
for thin films (b), $\mu_\text{add}$ is 210 cm$^2$V$^{-1}$s$^{-1}$ for $6\times10^{19}$ cm$^{-3}$, and 475 cm$^2$V$^{-1}$s$^{-1}$ for $7\times10^{19}$ cm$^{-3}$.}
  \label{fig:mobvsT-compexp}
\end{figure}

In Fig.~\ref{fig:mobvsT-compexp}(a) we show experimental values for bulk single crystals measured by Kim {\it et al.}\cite{kim-bso-mobility-2}
Their 300 K value at $8\times10^{19}$ cm$^{-3}$ doping, 320 cm$^2$V$^{-1}$s$^{-1}$,
is the highest RT mobility reported to date.
When comparing with experimental measurements it is important to recognize that
scattering mechanisms in addition to the LO-phonon and ionized impurity scattering may be present, for instance due to the presence of point defects such as compensating centers and neutral impurities, as well as extended defects such as dislocations or grain boundaries.
At large carrier densities ($>5\times10^{19}$ cm$^{-3}$), we find ionized impurity scattering to be temperature independent (see Fig.~\ref{fig:mobvsT-imp}); and neutral impurity scattering is also temperature independent\cite{Erginsoy1950}.
Therefore, we take these additional mechanisms into account via a {\em temperature-independent} mobility contribution $\mu_\text{add}$ chosen to reproduce the experimental {\em low-temperature} mobility value.

Taking these additional contributions into account, the calculated Hall mobilities for $8\times10^{19}$ cm$^{-3}$ and $1.2\times10^{20}$ cm$^{-3}$ doped samples agree very well with experiment.
For $8\times10^{19}$ cm$^{-3}$ doping, $\mu_\text{add}= 4500$ cm$^2$V$^{-1}$s$^{-1}$ indicating weak scattering due to other mechanisms.
However, for $1.2\times10^{20}$ cm$^{-3}$, the $\mu_\text{add}$ required for the fit was 1000 cm$^2$V$^{-1}$s$^{-1}$.
This suggests a significant presence of extended defects, neutral impurities or ionized defects.
Wang {\it et al.} have observed Ruddlesden-Popper type \{001\} stacking faults in BSO that could explain some of the additional scattering.\cite{Wang2015a}
However, more likely mechanisms, at such large dopant concentrations, include the formation of charge-compensating centers or carrier traps, or of neutral centers due to a decrease in the fraction of ionized dopants or because the dopant solubility limit is approached.
In Ref.~\onlinecite{kim-bso-mobility-2}, Kim {\it et al.} suggested that ionized impurity scattering was the dominant scattering mechanism in these samples, and neglected LO-phonon scattering in their analysis.
In contrast, as evident from our analysis related to Fig.~\ref{fig:mobvsden}(a) as well as Fig.~\ref{fig:mobvsT-compexp}(a), LO-phonon scattering that causes a temperature dependence plays an equally important role in limiting the RT mobility for $\sim10^{20}$ cm$^{-3}$ doping.

\subsubsection{Thin films}
BSO thin films\cite{kim-bso-mobility-2,Kim2013a,Kim2012,Schlom-Piper2016,Raghavan2016} exhibit a wide range in measured mobilities,
 which may reflect variations in growth technique and quality of the films.
The highest mobility reported is 150 cm$^2$V$^{-1}$s$^{-1}$ in thin films grown using molecular beam epitaxy\cite{Raghavan2016}.
In Fig.~\ref{fig:mobvsT-compexp}(b) we show an analysis similar to that used for bulk samples in Sec.~\ref{sec:compexp-bulk} to compare our calculated Hall mobility with the experimental values of Ref.~\onlinecite{Raghavan2016}.
Our analysis of the temperature dependence in thin films suggests strong additional scattering mechanisms to be present compared to bulk.
These additional scatterers could be related to dislocations or grain boundaries due to lattice mismatch with the substrate, as noted in the experimental reports\cite{Raghavan2016,kim-bso-mobility-2,Kim2013a}.
The temperature-independent contributions necessary to match the low-temperature mobility are $\mu_\text{add}$=210 cm$^2$V$^{-1}$s$^{-1}$ for $6\times10^{19}$ cm$^{-3}$, and $\mu_\text{add}$=475 cm$^2$V$^{-1}$s$^{-1}$ for $7\times10^{19}$ cm$^{-3}$.
We note that there seem to be additional {\em temperature-dependent} scattering mechanisms that reduce the mobilities more strongly with temperature than taken into account in our analysis.

One complication in thin films is its finite thickness (30--64 nm) in Ref.~\onlinecite{Raghavan2016}, which is smaller than or comparable to the average mean free path for electron-phonon scattering (calculated to be $v_{k{_\text{F}}}\langle\tau\rangle = 53$ nm).
This suggests that some of the carriers that have a momentum component perpendicular to the boundary will be limited by surface (or substrate-interface) scattering rather than LO phonon scattering\cite{ziman1960book}.
Therefore, in the calculation of mobility in thin films, inclusion of this effect will result in a reduction in mobility, and should be taken into account.
In fact, as reported in Ref.~\onlinecite{Raghavan2016}, increasing the film thickness from 30 nm to 64 nm increases the RT mobility from 100 to 124 cm$^2$V$^{-1}$s$^{-1}$.
Another finite-size effect is the close proximity to the substrate that could cause additional scattering from substrate phonons.

\subsection{Comparison to other perovskite oxides}\label{sec:compareSTO}

It is striking that the LO-phonon-limited mobility in BSO is about two orders of magnitude higher than in STO\cite{Himmetoglu2014,Verma2014a} and other perovskite oxides with CBs made up of $d$ states\cite{Sakai2009}.
We now show that this result provides some profound insights in the relative impact of various material properties on electron mobility.
The higher mobility of BSO has often been attributed to the lower effective mass of the CB.
This argument is based on the Drude model, where the mobility is given by $\mu = e\tau/m^*_\Gamma$, and assumes the scattering rate to be the same for both BSO and STO.
Here we point out that the scattering rates are actually significantly different in the two materials, and have a larger impact on the mobility than the effective masses.

Our calculated scattering rate in BSO, $\sim$10$^{13}$ s$^{-1}$, is an order of magnitude smaller than the scattering rate in STO, $\sim$10$^{14}$ s$^{-1}$, calculated in Ref.~\onlinecite{Himmetoglu2014}.
We now examine the origins of this difference. First, we direct our attention to the strength of electron-phonon coupling.
To quantify this strength, we calculated the value of the $q$-independent factor in the electron-phonon coupling matrix element $|g_{{ \bf q}\nu}|^2$ [Eq.~(\ref{eq:epmatrix})] for the second LO mode (LO$_2$), which we found to be the dominant scattering mode near RT (see Sec.~\ref{sec:LOphononresults} and Fig.~\ref{fig:mobvsT-LO}).
Using the LO and TO frequencies from Ref.~\onlinecite{Servoin1980}, this coupling factor in STO is computed to be 2.08$\times10^{-20}$ J$^2$m$^{-2}$.
This value is quite similar to the value computed for BSO, 1.77$\times$10$^{-20}$ J$^2$m$^{-2}$, and hence electron-phonon coupling strength cannot explain the difference in the scattering rate of BSO relative to STO.

The only other differences between BSO and STO that can explain the reduced scattering rate are the CB degeneracy and density of states (DOS).
In STO (and most other perovskite oxides) the CB is derived from $d$ orbitals of $t_{2g}$ character and hence is threefold degenerate near $\Gamma$ (possibly split by spin-orbit coupling, which is relatively weak in STO).
In contrast, BSO has a singly-degenerate CB composed of Sn $s$ orbitals.
This implies that in STO, each electronic state can scatter into two more bands via interband processes compared to BSO.
In addition, the lower CB dispersion in STO leads to an increase in the DOS.
Both these factors lead to an increase in the density of accessible final states in STO, which significantly raises the scattering rate.
We conclude that the higher mobility of BSO is not just due to a smaller effective mass, as previously emphasized\cite{Kim2012,mizoguchi-bso-opt-2}.
but more importantly to a significant reduction in the DOS that reduces the scattering rate.

This insight provides valuable guidance for selecting perovskite oxides for high-mobility applications.
The requirement of a small DOS does not necessarily mean that materials with CB derived from $d$ orbitals cannot exhibit high mobilities.
Any phenomena, such as spin-orbit coupling or biaxial strain, that remove the CB degeneracy by separating one or more bands away from the lowest CB by an amount exceeding the dominant LO phonon energy will reduce the scattering rate, and can enhance the mobility in perovskite oxides\cite{Himmetoglu2014}.

\subsection{Enhancing mobility in BaSnO$_3$} \label{sec:enhancemob}
From our discussion in Sec.~\ref{sec:mob-vs-den}, it is clear that ionized impurity scattering significantly limits the mobility in BSO, particularly at higher doping densities.
If high conductivity is required in bulk crystals or thick films, the presence of ionized impurities cannot be avoided since dopants are required for introducing electrons into the CB.
The problem of ionized impurity scattering has been recognized as the dominant mechanism limiting mobility at high doping in other TCOs as well\cite{tco-rev}.
In thin films, however, techniques are available to avoid or mitigate ionized impurity scattering.

One approach is to separate the ionized dopants from carriers in the channel;
this has been the basis of several semiconductor heterostructure designs\cite{Yu2005}.
A commonly used technique is modulation doping, where the dopants are introduced not in the channel, but in the barrier material.\cite{Yu2005}
Carriers from the dopants transfer into the channel forming a two-dimensional electron gas (2DEG), and are less prone to dopant scattering due to their spatial separation from the dopants.
BSO could benefit from such a doping technique provided appropriate barrier materials are identified\cite{Krishnaswamy2016}.

Perovskite oxides, including BSO, can also be doped by another technique, namely polar-discontinuity doping\cite{Ohtomo2004,Janotti2012,Moetakef2011,Bjaalie2014}.
This approach takes advantage of the polar discontinuity that exists, for particular interface orientations, between a nonpolar material (BSO, STO) and a polar perovskite oxide such as LaAlO$_3$, GdTiO$_3$, LaInO$_3$, or KTaO$_3$.
This polar discontinuity leads to the formation of an intrinsic 2DEG with a theoretical maximum density of 1/2 electron per interface atom ($\sim10^{14}$ cm$^{-2}$) in the nonpolar oxide, provided that the conduction-band offset is sufficiently high to confine the carriers.
Since the doping is intrinsic to the interface, carriers appear in the channel without the need for extrinsic doping.
The host atoms in the interfacial layer effectively act as donors, but since they are arranged in a periodic lattice they do not give rise to scattering.
This eliminates ionized impurity scattering, and thus in principle offers mobilities close to the phonon-scattering-limited value.

STO has been the material of choice to explore polar-discontinuity doping\cite{Moetakef2011,Ohtomo2004} as well as other functional properties\cite{Ismail-Beigi2015} in perovskite oxides, but it suffers from a low RT mobility of $\sim$10 cm$^{2}$V$^{-1}$s$^{-1}$.\cite{Himmetoglu2014,Verma2014a}
BSO, which has an intrinsic RT mobility in the range of 300--600 cm$^{2}$V$^{-1}$s$^{-1}$ [Table~\ref{table:mobility}] thus presents an interesting high-mobility alternative to STO.
However, the low DOS of BSO makes confining the 2DEG within the channel challenging; a careful design of the heterostructure barriers with a large enough CB offset is necessary.
Issues related to confinement and DOS have been quantitatively addressed in Ref.~\onlinecite{Krishnaswamy2016}, along with design guidelines for BSO heterostructures based on modulation doping and polar-discontinuity doping.

\section{Conclusion} \label{sec:conclusion}

In summary, we have used first-principles analysis along with a careful numerical procedure to calculate the mobility of BSO from Boltzmann transport theory within the relaxation time approximation, accounting for LO-phonon scattering as well as ionized impurity scattering.
We find that the surprisingly large mobility of BSO stems not only from the small effective mass (as had been previously suggested), but is also due to a significant reduction in the LO-phonon scattering rate compared to other perovskite oxides.
The reduction in the rate is shown to be due to a decrease in the number of states that the electrons can scatter into because of the low DOS in BSO.
Ionized impurity scattering was found to be a significant scattering mechanism, even at RT, limiting the mobilities to less than 330 cm$^2$V$^{-1}$s$^{-1}$ for dopant densities above $10^{19}$ cm$^{-3}$.
Ionized impurity scattering can be avoided by using modulation doping or polar-discontinuity doping, which may enable achieving LO-phonon-limited mobility values, which are calculated to exceed 500 cm$^{2}$V$^{-1}$s$^{-1}$ for electron concentrations $>1\times10^{19}$ cm$^{-3}$.

\section*{Acknowledgements}
We are grateful to Prof. S. James Allen for fruitful discussions.
This work was supported by the Center for Low Energy Systems Technology (LEAST), one of six SRC STARnet Centers sponsored by MARCO and DARPA, and
by the MURI program of the Office of Naval Research, Grant No. N00014-12-1-0976.
Computational resources were provided by the Center for Scientific Computing at the CNSI and MRL (an NSF MRSEC, DMR-1121053) (NSF CNS-0960316), and by the Extreme Science and Engineering Discovery Environment (XSEDE), supported by NSF (ACI-1053575).

\end{document}